\documentclass[intlimits,twoside,a4paper]{article}
\usepackage[utf8]{inputenc}
\usepackage[eqsecnum]{cmpj3}

\usepackage{xcolor}
\usepackage[mathscr]{eucal}
\usepackage{bm}

\issue{2022}{25}{3}{33603}
\doinumber{10.5488/CMP.25.33603}

\title[Contact angle from a density functional theory]
{Contact angle of water on a model heterogeneous surface. A density functional approach
}

\author[K. D\k{a}browska, O. Pizio, S. Sokołowski]
{K. Dąbrowska\refaddr{label1},
O. Pizio\orcid{0000-0001-8333-4652}\refaddr{label2},
S. Sokołowski\orcid{0000-0003-0580-5214}\refaddr{label1}}

\addresses{
\addr{label1} 
Department of Theoretical Chemistry,
Maria Curie-Sk{\l}odowska University,  Lublin 20-031, Poland,\\ email: stefan.sokolowski@gmail.com
\addr{label2}
Instituto de Qu\'{i}mica, Universidad Nacional Aut\'{o}noma de M\'{e}xico,
Circuito Exterior, 04510, Cd. de M\'{e}xico, M\'{e}xico, email: oapizio@gmail.com}

\date{Received May 11, 2022, in final form July 13, 2022}

\begin{document}
\maketitle

\begin{abstract}
We use a density functional approach to calculate the contact angle of the water model on a
heterogeneous, graphite-like surface. 
The surface heterogeneity results from the pre-adsorption of a layer of spherical 
species. The pre-adsorbed molecules can also be a mixture of molecules of different sizes.
The presence of pre-adsorbed layer causes geometrical and energetical heterogeneity of the surfaces.
Two cases are considered. The pre-adsorbed molecules can either behave like hard-sphere obstacles, or they can also attract the molecules of water. In the first case, an increase of the amount of 
pre-adsorbed species leads to an increase of the wetting temperature, but 
this increase does not depend linearly  on the amount of obstacles. 
In the case of obstacles exerting attractive forces on water molecules, 
the curves describing the dependence between the amount of pre-adsorbed 
species and the contact angle  can exhibit a maximum. 
In addition, we have also studied how the pre-adsorbed species influence 
the local densities of gaseous and liquid phases in contact with a modified solid surface.

\keywords
density functional theory, contact angle, heterogeneous surface, water model


\end{abstract}

\section{Introduction}

The problem of the influence of surface heterogeneity of 
solid surfaces on thermodynamic properties of fluid-solid systems
 has been the subject of intensive research for a long time~\cite{sips,house,baka,rududu,xia,kumar,li,barrow}.
Usual models of heterogeneous surfaces have been based on distinguishing
several kinds of surface active sites. The sites differ by their
interaction energy with fluid molecules.  
The function describing the distribution of sites with the energy is termed
``the energy distribution function'' and
 is commonly treated as a  characteristic property of a given heterogeneous solid~\cite{rududu}.
 
Different models describing the systems with heterogeneous surfaces have been considered.  They can be classified according to the way in which the surface active sites are  defined and distributed over a surface.
According to the so-called patchwise model, the sites
of the same kind are grouped into patches. The patches are geometrically flat,
energetically homogeneous and  
 independent of each other.   
Consequently,  thermodynamic properties of a fluid in 
contact with a patchy heterogeneous solid (e.g., adsorption isotherms) are computed as 
weighted averages of the relevant properties for homogeneous surfaces 
with statistical weights 
resulting from the energy distribution function~\cite{house,rududu}.
However, for several solid surfaces, such as silica gels, aluminum oxides, or active
carbons, the patchwise heterogeneous model is unrealistic~\cite{kumar} and considering
such systems, the models of random heterogeneous surfaces were
introduced~\cite{ali,adamczyk}.

One of the possible statistical-thermodynamic methods of describing the random
heterogeneous systems comes from the theory of 
quenched-annealed systems~\cite{pizio,cuesta,schmidt}. Although the latter approach is mainly applied in
the studies of fluids confined by microporous solids, it can be equally well used for modelling a rough
solid surface, or capillaries filled with random matrices~\cite{dong,nasze1,nasze2}.
The initial flat surface is covered with a layer
of randomly distributed particles, which remain frozen at fixed positions
when the fluid enters the system.

The models intermediate between patchwise and random surfaces
are patterned surface models. The sites of particular kinds
occupy small regions of simple
geometrical shape (e.g., stripes, squares, or triangles). These geometrical
constructs are ordered into crystalline-like lattices.
In contrast to patchwise models, 
the processes occurring over particular structures
are not independent of each other.
The thermodynamic properties of fluids in contact with  
patterned surfaces were studied in several publications, see e.g.
\cite{patt1,patt2,patt3,patt4,pa}.

The heterogeneity of solid surfaces plays an important role 
in the adsorption of fluids on solids and in 
chromatographic processes~\cite{chroma1,chroma2}. Indeed, 
the presence and  specific distribution of adsorption
sites on the surfaces can decide about the effectiveness
of the separation of fluid mixtures.
Furthermore, the surface heterogeneity can strongly influence the surface phase transitions.
The studies of the latter problems mainly concern patterned and random surfaces~\cite{dft0,dft1,dft2,dft3}.

Wetting is an important phenomenon that is common in nature~\cite{bonn}.
It is also an essential issue for the design and development 
of novel substances with desirable surface properties that
can be applied in many areas of engineering,  chemistry, and biology. 
One of the most significant ingredients that appear in several processes is water.
In particular, the wetting behavior of water on graphite-like substrates was 
the subject of our recent detailed study~\cite{nasza4}.

Experimental method for investigating
the wetting is usually based on the measurement of the static contact angle, $\theta$. 
A crossover from non-wetting to wetting state takes place if
the contact angle changes from a nonzero value 
to zero~\cite{Dietrich88}.
The temperature at which this transition occurs is called the wetting temperature, $T_w$.
Other experiments 
rely on the measurements of adsorption isotherms for gas densities up to 
the saturated vapor density.   
In the case of first-order wetting transitions, characteristic
changes in the course of adsorption isotherms with temperature take place.
Namely, at temperatures below $T_w$,  upon the bulk density approaching
the liquid-vapor coexistence, the thickness of the adsorbed film
remains finite and small,  whereas at higher temperatures, it diverges to infinity~\cite{evans3,evans,ev}.
Thus, the study of changes of adsorption isotherms allows the determination of $T_w$. 
Note that the adsorption method for the first time permitted to capture the wetting temperature 
for ${^4}{\rm He}$ on cesium~\cite{taborek}.

Of course, the value of the static contact angle, $\theta$, depends on 
the surface heterogeneity. However, there is no general approach that
would describe $\theta$ in terms of parameters characterizing the surface
heterogeneity. 
Although several attempts to elucidate the 
effects of surface heterogeneity on the values of $\theta$ were undertaken,
but they concerned specific surface models.
For patchy heterogeneous surfaces, 
Cassie and Baxter~\cite{cassie}
proposed an expression relating the contact angle to the energy distribution function.
Another attempt was proposed by Wenzel~\cite{wenzel} for macroscopically rough surfaces.
Both Cassie-Baxter and Wenzel  approaches were next  applied
and verified for numerous systems, cf. reference~\cite{erb}
and the references quoted therein. 
The Cassie-Baxter and
Wenzel equations can be used if the patches are big enough, 
i.e., if their size is much larger than the range of capillary forces. 
We would like to stress that the Cassie-Baxter equation for
the contact angle is familiar~\cite{lipovsky} to the so-called integral equation 
for the adsorption isotherm,~\cite{rududu}.
Recent studies, however,  concentrated on the description of patterned surfaces
composed of alternately arranged
 hydrophilic and hydrophobic regions~\cite{s1,s2,s3,s4,s5,s5a,s5b}.

One of the methods~\cite{i1,i2} for modifying the surface properties of solids relies on  pre-adsorption of some selected species on bare surfaces, e.g.,
chain molecules (oligomers and polymers).
This method has found particularly important applications in developing novel 
chromatographic column packing with the required properties~\cite{j1,j2}.
The grafted chains 
have also a significant impact on the surface phase transitions,
such as wetting, layering, and capillary condensation
in confined fluids~\cite{ch1,ch2,ch3}.  As our recent study indicated, the grafted
chains are capable not only of quantitatively but also qualititavely influencing
the topology of the surface phase diagrams~\cite{ch4}.

Theoretical methods for the description of surface phase transitions
are main\-ly based on
the density functional approaches~\cite{wandelt}.
Density functional theories can be also employed to the systems
with surfaces modified by chemically bonded spherical molecules. Therefore,
in this work, we propose a density functional method to study the
changes of static contact angle with the changes  of the adsorbing  surface
amount and size of pre-adsorbed spherical molecules.

The fluid-fluid
interaction is selected to mimic the interaction between a pair
of water molecules.
Water belongs to the class of associating fluids and
the formation of intermolecular hydrogen (associative)
bonds should be taken into account in the model. However, 
the application of contemporary force fields for water 
with site-site electrostatic forces~\cite{omar17} would be
computationally prohibitive within the density functional approaches.  
Therefore, we use the potential resulting from
the statistical association fluid theory (SAFT),
\cite{gubbins2,gubbins3} with the parameters proposed
by Clark et al.~\cite{clark}. This model
 is not only one of the most accurate in terms 
of evaluating the liquid-vapor phase diagram, but it also correctly predicts the temperature dependence of the surface tension~\cite{clark,gloor004}.

Water molecules are in contact with a modified surface of graphite. Similarly
to our previous work~\cite{nasza4}, the Lennard-Jones 10-4-3 function~\cite{steele} is
used to describe the interaction of a water particle with a bare graphite surface. 
The surface of graphite
is covered with a layer of pre-adsorbed spherical molecules. The pre-adsorbed layer can be either one- or many-component and we consider
the cases of purely repulsive and repulsive and attractive interactions with water
molecules.
The values of the equilibrium contact angle are then calculated
from the density functional theory. The theory used for us is a modification
of the approach for grafted chain molecules~\cite{nasze1,nasze2}. The pre-adsorbed layer 
creates both geometrical and energetic 
heterogeneity of an adsorbing surface.
However, we should also mention here that there exist
alternative density functional approaches proposed by Aslymov and co-workers~\cite{asly,asly1}
and by Zhou~\cite{zhoo},
based on the development of an appropriate expression for the
``effective'' fluid-solid that takes
into accounts the  surface roughness. This effective potential is then
used as an external potential field in the classical density functional
treatment. Our approach, however, leads to modifications of
the expressions for the system free energy.

The paper is arranged as follows. In the next section we briefly outline the details of the model and the interaction potentials. Then, we describe
the basic points of the density functional theory and the method for 
the evaluation of the values of the contact angle. Section 3 presents the 
results obtained for the pre-adsorbed layer formed by one-component
layer of hard spheres (obstacles), built of a binary mixture of
obstacles, and, finally, we consider the case of pre-adsorbed particles
interacting via repulsive and attractive forces. The last section concludes
the obtained theoretical data.

\section{Model and theory}\label{model} 

\subsection{Interaction potentials}

According to the model developed by Chapman, Gubbins, and Jackson~\cite{jackson1,jackson2},  
each water molecule possesses four associative sites $\Gamma=\{A, B, C, D\}$, located
at the vertices of tetrahedron inscribed into a spherical core.
The interaction energy between the molecules $i=1,2$ depends
on the center-to-center distance, $r_{12}=|\mathbf{r}_{12}|$ ,
and on orientations of both molecules, ${\bm \omega}_i$,
\begin{equation}
 u(12) = u_{ff}(r_{12}) + \sum_{\alpha \in \Gamma} \sum_{\beta \in \Gamma} 
u_{\alpha\beta}(\mathbf{r}_{\alpha\beta}).
\end{equation}
The site-site vectors,  $\mathbf{r}_{\alpha\beta}$, are
$\mathbf{r}_{\alpha\beta}=\mathbf{r}_{12}+\mathbf{d}_{\alpha}(\bm{\omega_1)}-\mathbf{d}_{\beta}(\bm{\omega_}2)$,
where $\mathbf{d}_{\iota}(\bm{\omega_}i)$
is the vector connecting the site $\iota$ on molecule $i$ with its center (see also figure~1 of reference~\cite{jackson1}).
Only the site-site association  AC,  BC, AD, and BD is allowed, and
all association energies are  equal. The associative interaction between the sites
is 
\begin{equation}
\label{eq:asw}
u_{\alpha\beta}(\mathbf{r}_{\alpha\beta})=
\left\{
\begin{array}{ll}
-\varepsilon_{{\rm as}}, &    0< |\mathbf{r}_{\alpha\beta}| \leqslant r_c ,\\
0,  &     |\mathbf{r}_{\alpha\beta}|   > r_c ,
\end{array}
\right. 
\end{equation}
where $\varepsilon_{{\rm as}}$ is the depth  and
$r_c$ is the cut-off range of the associative interaction. 

The non-associative part of the pair potential, $u_{ff}(r)$, is described by a square-well potential
\begin{equation}
u_{ff}(r) = u_{{\rm hs},ff}(r) +  u_{{\rm att},ff}(r),
 \label{eq:sw}
\end{equation}
where $u_{{\rm hs},ff}(r)$ and $u_{{\rm att},ff}(r)$ are, respectively, the
hard-sphere (hs) and attractive (att) parts of the potential, 
\begin{equation}
u_{{\rm hs},ff}(r) =
\left\{
\begin{array}{ll}
\infty, & r < \sigma,\\
0, & r \geqslant \sigma,
\end{array}
\right.
\end{equation}
and
\begin{equation}\label{uSW}
 u_{{\rm att},ff}(r)=
\left\{
\begin{array}{ll}
0, & r < \sigma,\\
\varepsilon, & \sigma \leqslant r < \lambda_{ff} \sigma,  \\ 
0, & r \geqslant \lambda_{ff} \sigma.
\end{array}
\right.
\end{equation}
In the above,
$\sigma$, $\varepsilon$ and $\lambda_{ff}$ are the diameter, the depth and the range of the
non-associative water-water potential, respectively.

The parameters for
the W1 model of Clark et al.~\cite{clark}  are collected in table \ref{tab:1}.
\begin{table}[h] 
  \centering
   \caption{
  The parameters of the W1 water-water model potential
   from reference~\cite{clark}.
   \vspace{0.1cm}
   }
   \begin{tabular}{ccccccc
   }
  \hline \hline
 Model &  $\sigma  $ (nm)& $\varepsilon/k$ ($K$) & $\lambda$&  $r_c $ (nm)&
     \vspace{0.1cm} $\varepsilon_{{\rm as}}/k$ ($K$) & $|{\mathbf{d}}_\iota|/\sigma$  \\[0.5ex]
     \hline
W1 & 0.303420 & 250.000 & 1.78890 & 0.210822 & 1400.00 & 0.25 \\
\hline \hline
\end{tabular}
\label{tab:1}
\end{table}

Water molecules are in contact with a modified surface.
The initial, non-modified surface is assumed to be graphite-like.
The adsorbing potential, $v(z)$, exerted  on a fluid particle 
is then given by the potential developed by Steele~\cite{steele}.
\begin{equation}
 v(z) =\varepsilon_{fw}
 \left[ \frac{2}{5} \left( \frac{ \sigma_{fw}}{z}\right)^{10} 
 - \left( \frac{ \sigma_{fw}}{z}\right)^{4} \right. 
\left. - \frac{\sigma_{fw}^4 }
{3 \Delta (z+0.61 \Delta)^3 }
 \right],
 \label{eq:steele}
\end{equation}
where $\varepsilon_{fw}$ and  $\sigma_{fw}$ are
the energy and the size parameters,  respectively. The interlayer spacing
of the graphite planes equals $\Delta= 0.335$~nm. The values of  $\sigma_{fw}$ 
and $\varepsilon_{fw}$ follow 
 from Lorentz-Berthelot mixing  rules. Thus,~\cite{nasza4}, $\sigma_{fw}/\sigma=1.06028$
 and $\varepsilon_{fw}/\varepsilon=8.311$. Our calculations were also
 carried out for $\varepsilon_{fw}/\varepsilon=7$ and 9.5. 
 The potential of Steele
was widely used with success in the theory of adsorption of fluids on graphite~\cite{nasza4,thesis}.
A comprehensive discussion of this potential was provided  by Zhao and Johnson~\cite{Zhao2005}.

The inhomogeneity of the surface results from its chemical modification. 
According to our model, the modification means the  ``sticking'' (tethering)
of spherical particles of diameters $\sigma_i$
at the  distance $\sigma_i/2$ from the surface.
For the surface at $z=0$, the potential that leads to the tethering of molecules reads
\begin{equation} 
\label{eq:delta}
\exp[-v_{i}^{(1)}(z)/kT]= \delta (z-\sigma_{i}/2),
\end{equation}
where $\delta$ denotes the Dirac function.

The amount of tethered species $i$ is controlled by the parameter $R_{ci}$ that gives
the total surface density of pre-adsorbed species. (In the case of one-component
pre-adsorbed layer, we drop the subscript $i$ and use the symbol $R_c$.)

Similarly to the case of non-associative water-water interaction, the interaction of
water molecules with pre-adsorbed species,
$u_{fi}(r) = u_{{\rm hs},fi}(r) +  u_{{\rm att},fi}(r)$,
is described by the square well potential
\begin{equation}
u_{{\rm hs},fi}(r) =
\left\{
\begin{array}{ll}
\infty, & r < \sigma_{fi},\\
0, & r\geqslant \sigma_{fi},
\end{array}
\right.
\end{equation}
and
\begin{equation}\label{uSW1}
 u_{{\rm att},fi}(r)=
\left\{
\begin{array}{ll}
0, & r < \sigma_{fi},\\
\varepsilon_{fi}, & \sigma_{fi} \leqslant r < \lambda_{fi} \sigma_{fi},  \\ 
0, & r \geqslant \lambda_{fi} \sigma_{fi}.
\end{array}
\right.
\end{equation}
The interactions between pre-adsorbed species, however, are assumed to be of hard-sphere type
with the cross size parameters equal to $(\sigma_i+\sigma_j)/2$.

\subsection{Density functional theory}

The system is studied using a version of the density functional theory (DF),
described already in detail in references~\cite{nasze1,nasze2}. To avoid unnecessary
repetitions, we recall only basic equations. 

 The symbols  
 $\rho(\mathbf{r})$ and  $\rho_i(\mathbf{r})$ denote
the local density of water-like molecules and of tethered particles of the kind $i$. 
As usual, we start with defining the excess surface free energy (the grand canonical potential)
as a functional of the local densities
\begin{equation}
 {\Omega} = F[\rho(\mathbf{r}),\{\rho_i(\mathbf{r})\}] + \sum_{\{i\}}\int \rd\mathbf{r} \rho_i v_i(z) +  
 \int \rd \mathbf{r} \rho(\mathbf{r})[v(z)-\mu],
 \label{eq:omega}
\end{equation}
where $\mu$ is the chemical potential of water and  $F[\rho(\mathbf{r}),\{\rho_i(\mathbf{r})\}]$.
At fixed values of temperature and the amount of pre-adsorbed species of all kinds, $\Omega$ is a function
of $\mu$, or, alternatively, the bulk density of water, $\rho_b$, $\Omega=\Omega(\rho_b)$.

The free energy functional is considered as the sum of an ideal term, 
$F_{{\rm id}}[\rho(\mathbf{r}),\{\rho_i(\mathbf{r})\}]$,
the hard-sphere functional
$F_{{\rm hs}}[\rho(\mathbf{r}),\{\rho_i(\mathbf{r})\}]$, the functional arising from attractive interparticle forces, 
$F_{{\rm att}}[\rho(\mathbf{r}),\{\rho_i(\mathbf{r})\}]$ and the part
due to the
formation of associative bonds, 
$F_{{\rm as}}[\rho(\mathbf{r}),\{\rho_i(\mathbf{r})\}]$.
\begin{align} 
 F[\rho(\mathbf{r}),\{\rho_i(\mathbf{r})\}]&=
 F_{{\rm id}}[\rho(\mathbf{r})] 
 F_{{\rm hs}}[\rho(\mathbf{r}),\{\rho_i(\mathbf{r})\}] \nonumber \\
 &+F_{{\rm att}}[\rho(\mathbf{r}),\{\rho_i(\mathbf{r})\}]+
 F_{{\rm as}}[\rho(\mathbf{r}),\{\rho_i(\mathbf{r})\}].
\end{align}
The ideal term is known exactly
\begin{equation}
 F_{{\rm id}}/kT=
 \int \rd \mathbf{r} \rho(\mathbf{r}) [\ln\rho (\mathbf{r})-1].
\end{equation}
The hard-sphere functional is evaluated according to
the theory originally developed by Rosenfeld and modified in~\cite{61}.
It requires the knowledge of four scalar [$n^{(I)}(\mathbf{r})$, $I=0,1,2,3$]  and two vector
[$n^{(VI)}(\mathbf{r})$, $I=1,2$] averaged densities
\begin{equation}
 n^{(L)}(\mathbf{r})=\int  \rd\mathbf{r}' \rho(\mathbf{r}') w^{(L)}(|\mathbf{r}-\mathbf{r}'|)+
\sum_{\{i\}} \int  \rd\mathbf{r}'  \rho_i ({\bm r'}) w^{(L)}_i(|\mathbf{r}-\mathbf{r}'|),
\end{equation}
where $L=0,1,2,3,V1,V2$ and $ w^{(L)}(|\mathbf{r}-\mathbf{r}'|)$ are the weight functions, see equations~(11)--(14) of
reference~\cite{61}.
However, the explicit equations for the hard-sphere free energy is given by equation~(11) of~\cite{nasze1}.

The attractive forces contribution
results from the mean-field approximation. According to the introduced model, the pre-adsorbed 
molecules do not interact with each other by attractive forces. Thus,
\begin{align}
 F_{{\rm att}}[\rho(\mathbf{r}),\{\rho_i(\mathbf{r})\}]&= \frac{1}{2}
 \int\int  \rd\mathbf{r} \rd\mathbf{r}'\rho(\mathbf{r}')\rho(\mathbf{r})u_{{\rm att},ff}(|\mathbf{r}'-\mathbf{r} |) \nonumber \\ 
 &+
 \sum_{\{i\}} \int\int  \rd\mathbf{r} \rd\mathbf{r}'\rho(\mathbf{r}')\rho_i (\mathbf{r})u_{{\rm att},fi}(|\mathbf{r}'-\mathbf{r} |).
\end{align}
Finally, the contribution arising from the formation of associative bonds
between water molecules  follows from the application 
of the first-order thermodynamic
theory of  Wertheim (TPT1). 
The TPT1 contribution to the free energy is expressed in terms of
the fraction of  molecules not bonded at a given site $\chi(\mathbf{r})$, 
which represents the statistical mechanical analogue of the mass action law.
The expression used by us for $F_{{\rm as}}[\rho(\mathbf{r}),\{\rho_i(\mathbf{r})\}]$ reads~\cite{ch3}
\begin{equation}
F_{{\rm as}}[\rho]= 4 \int \rd\mathbf{r} \,  n_0(\mathbf{r}) \zeta(\mathbf{r})
\left\{\ln\chi(\mathbf{r})-\frac{1}{2} \left[\chi(\mathbf{r})-1 \right] \right\}. 
\end{equation}
All the details and the definition of the functions $\zeta(\mathbf{r})$  and $\chi(\mathbf{r})$ 
are given in previous works~\cite{ch3,trejos3}. We omit here their explicit
expressions to avoid unnecessary repetitions.

The density profile $\rho(\mathbf{r})$ results from the Euler-Lagrange equation, 
\begin{equation}
\delta \Omega / \delta \rho(\mathbf{r})=0.
\end{equation}
The density profiles of pre-adsorbed species, however, are fixed by the 
external potential, equation~(\ref{eq:delta}), and the assumption about constancy
of pre-adsorbed particles of the kind $i$. 
This means that
\begin{equation}
 \int \rd z \rho_{i}(z)=R_{ci},
\end{equation}
i.e., $\rho_i(z)=R_{ci}\delta(z-\sigma_i/2)$.

If the external field depends on the 
distance perpendicular to the solid surface only, as in equation~(\ref{eq:steele}), 
then the local density is one-dimensional as well, i.e. $\rho(\mathbf{r})\equiv\rho(z)$.
The Euler-Lagrange equation for the density profile
can be solved using different numerical procedures, see e.g.,~\cite{roth1} and references
therein. In our calculations, we used the classical Picard iteration method.

\subsection{Contact angle}

The criterion for wetting the surface by a liquid  is usually derived from the classical
Young's equation that expresses the force balance at a three-phase
contact between a liquid drop ($l$), a solid  surface ($s$) and a gas
phase ($g$), in terms of the contact angle, $\theta$~\cite{Dietrich88,evans3} as
\begin{equation}
 \gamma_{g}-\gamma_{l}=\gamma \cos\theta,
 \label{young}
\end{equation}
where $\gamma_{\kappa}=\Delta \Omega_{\kappa}/A$ is the interfacial tension
at gas-solid ($\kappa=g$) and at liquid-solid ($\kappa=l$) interface, $A$ is the
interfacial surface area,
$\Delta\Omega{_g}=\Omega(\rho_{bg})-\Omega_{b}(\rho_{bg})$ and 
$\Delta\Omega{_l}=\Omega(\rho_{bl})-\Omega_{b}(\rho_{bl})$ are the
excess grand thermodynamic potentials for the  gas-solid
and for the liquid-solid interfaces calculated for the bulk densities
$\rho_{bg}$ and $\rho_{bl}$ corresponding to the densities of coexisting
gaseous and liquid bulk phases.
Next, $\Omega_b$ is the bulk grand thermodynamic
potential at the bulk liquid-vapor coexistence and $\gamma$ is the gas-liquid interfacial
tension (the surface tension).

The Young equation is an approximation to reality~\cite{jasper}
as it neglects the effect of the line tension at a three-phase
contact.
The wetting temperature, $T_w$, is determined as the highest temperature
at which the ratio $(\gamma_{g}-\gamma_{l})/\gamma$ becomes zero
for the first time.
At temperatures $T\geqslant T_w$,
the liquid drop completely spreads across the surface.

The applications of equation~(\ref{young})
requires the knowledge 
of  the values of the surface tension,  $\gamma$.
In order to evaluate them, 
one needs to calculate the changes of the local density across the interface between 
semi-infinite slab of a liquid and a semi-infinite slab of a gas.   
This is  done by removing the solid
wall and setting the boundary conditions
to $\rho(z=-\infty) = \rho_l$ and  $\rho(z=\infty)=\rho_g$, using the procedure
described in detail in reference~\cite{bryk004}.

\begin{figure}[!h]
	\begin{center}
	\includegraphics[scale=0.4]{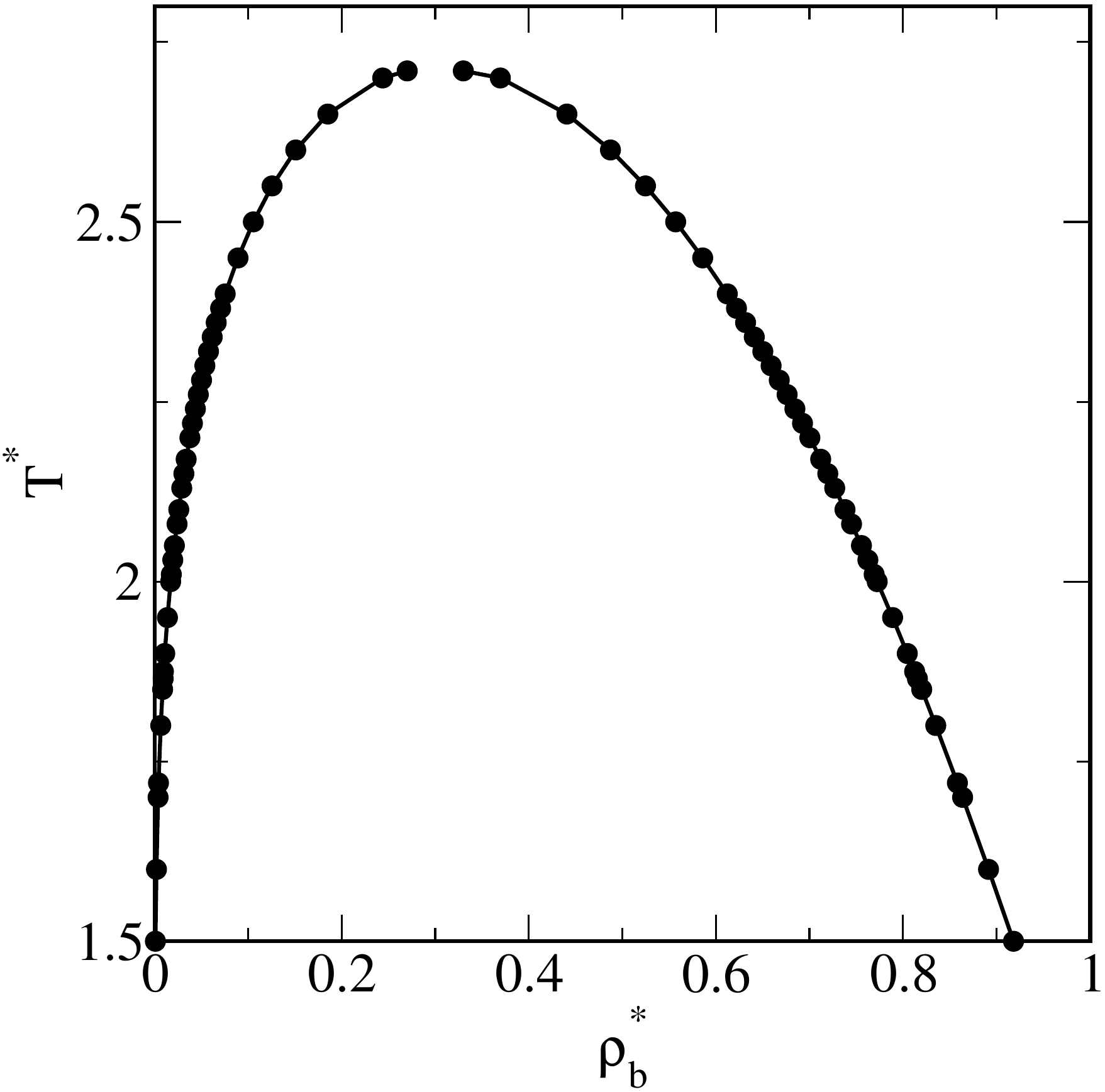}
	\end{center}
	\vspace{-0.8cm}
	\caption{ Bulk liquid-vapor phase diagram in the temperature-density plane for the model W1 of water.} 
	\label{fig:0}
\end{figure}

\section{Results and discussion}

All variables below are expressed in  reduced units.
The reduced temperature, reduced distance,
reduced diameter and a reduced amount of pre-adsorbed particles, as well as
a reduced  local and bulk density are $T^* = kT/\varepsilon$,
$z^* = z/\sigma$, $\sigma_i^*=\sigma_i/\sigma$, $R_c^*=R_c\sigma^2$ and
$\rho^*(z)=\rho(z)\sigma^3$, and
$\rho_b^* = \rho_b\sigma^3$, respectively.

Evaluation of the contact angles 
requires the knowledge of two bulk densities  at the gas-liquid coexistence.
The calculation of the bulk phase diagram was self-consistently carried out 
using the bulk version of the density functional theory~\cite{nasza4}, and
the obtained phase diagram is given in figure~\ref{fig:0}. Similarly, the values of the liquid-vapor
surface tension were obtained using the density functional 
method and all the details of these calculations are outlined in~\cite{nasza4}.

We have already noted that similarly to  our previous works~\cite{nasza4,ch4}, the calculations were
carried out for the W1 model of Clark et al.~\cite{clark}.
The parameters of this model  were selected so as
to reproduce the experimental bulk  
liquid-gas phase diagram. Therefore, the agreement between  theoretical and
experimental bulk dew and bubble 
densities is quite good, in general. However, the  data fitting  of Clark was at
temperatures lower than the bulk critical temperature. Thus, some deviations can appear
between bulk theoretical and experimental data
within the temperatures range in the vicinity of the critical temperature.
When assessing the results presented below, we should remember that a lower accuracy 
of the bulk system description  for temperatures close to 
the critical temperature may have an impact on the predictions of the applied theory.
Note that the bulk critical temperature resulting from our approach ~\cite{nasza4,trejos1},
$T_c^*\approx 2.715$, is in a reasonable agreement with experiment.

\begin{figure}[htb]
	\begin{center}
	\includegraphics[width=6.9cm]{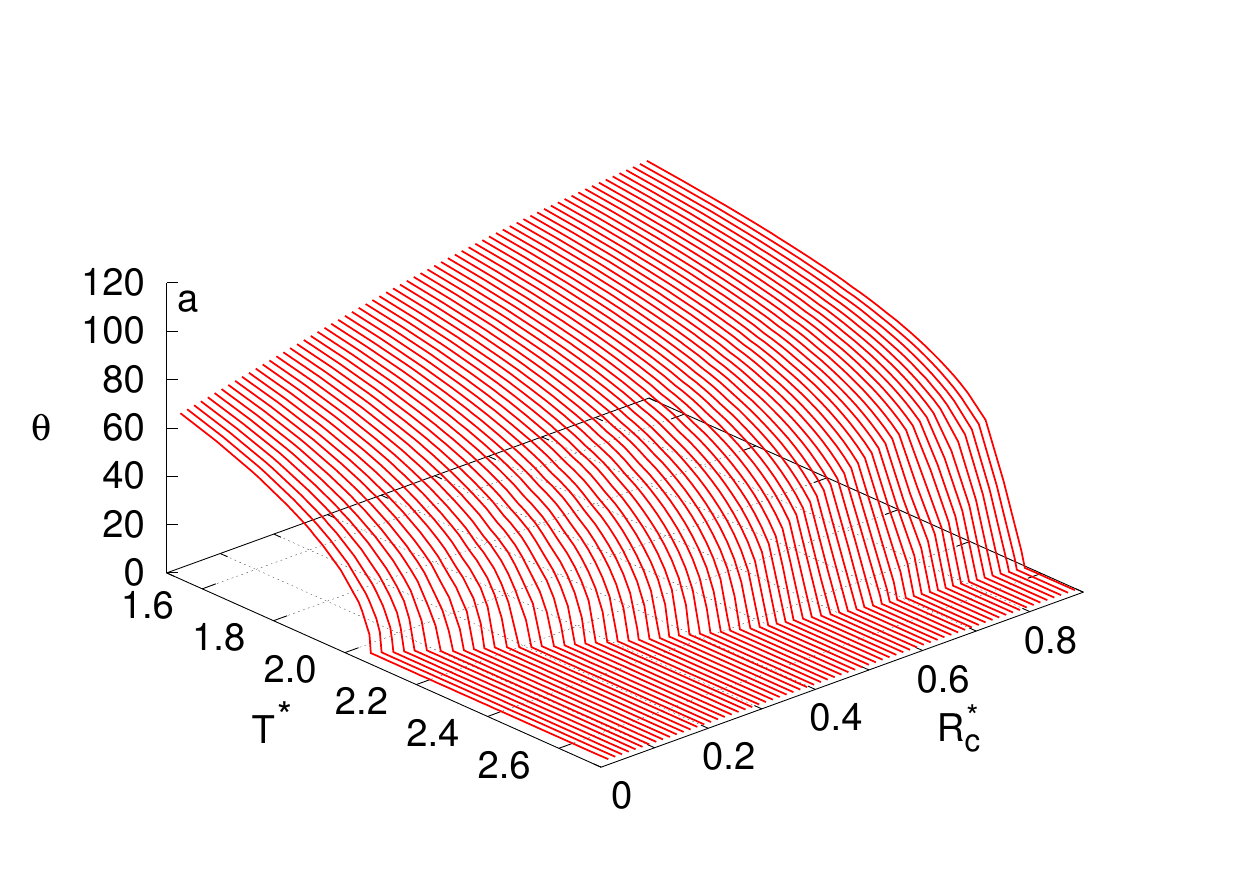}
	\includegraphics[width=6.9cm]{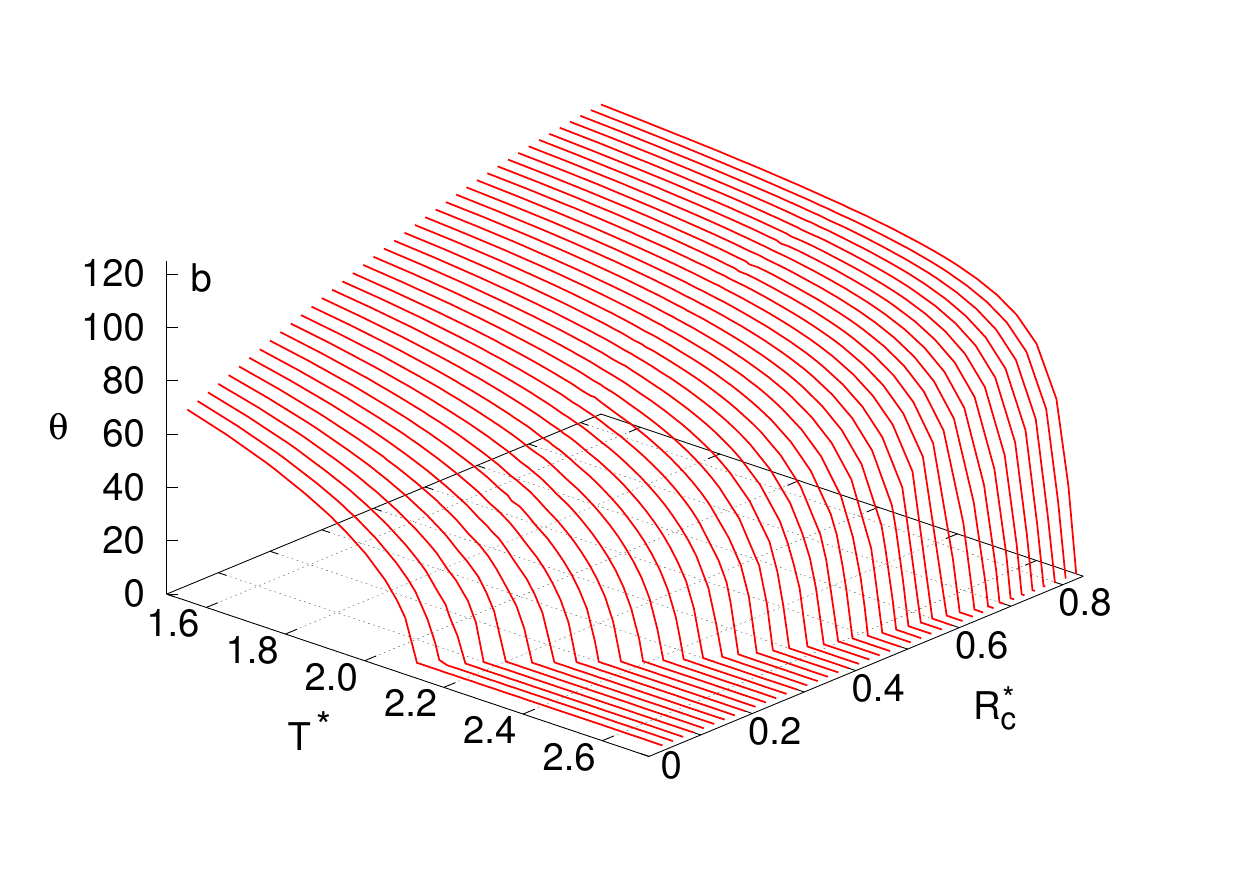}
	\includegraphics[width=6.9cm]{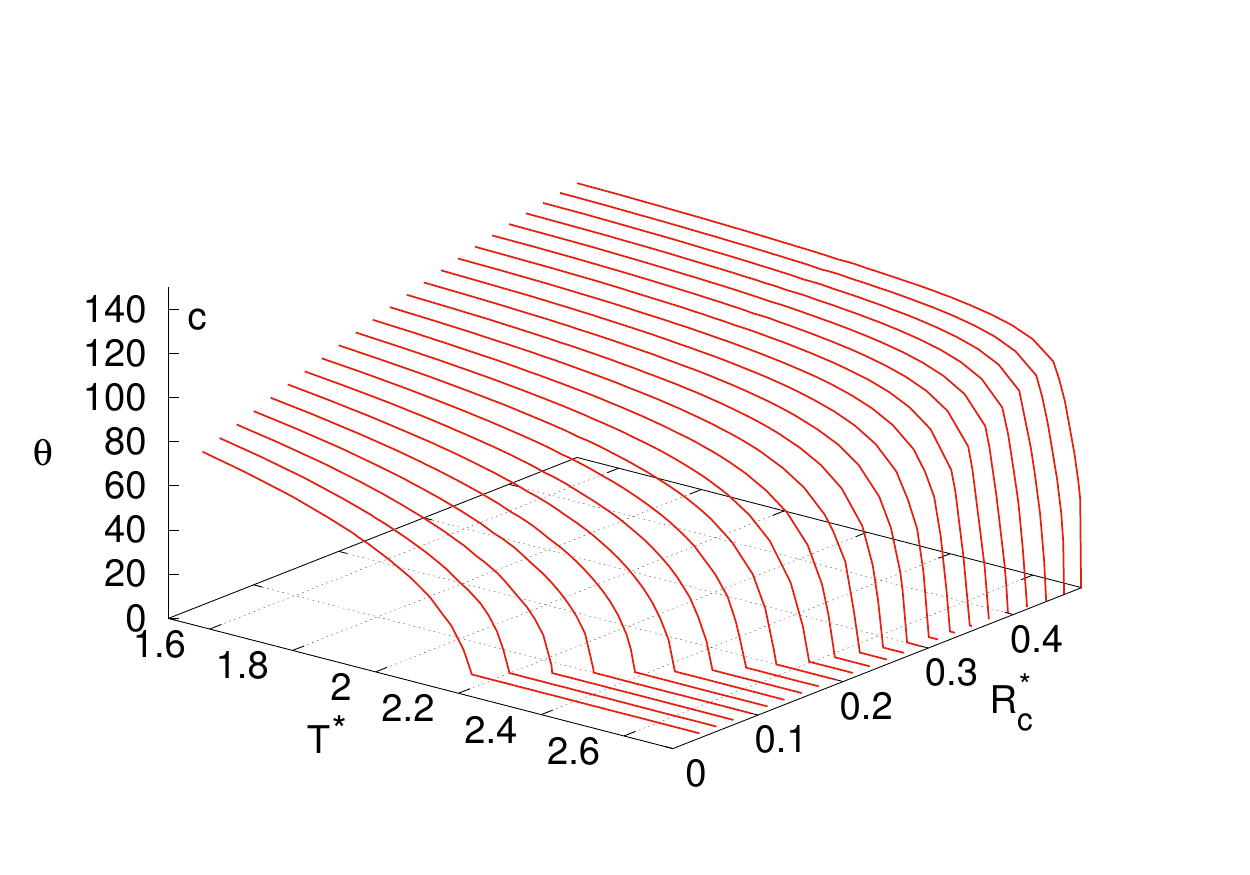}
	\includegraphics[width=6.9cm]{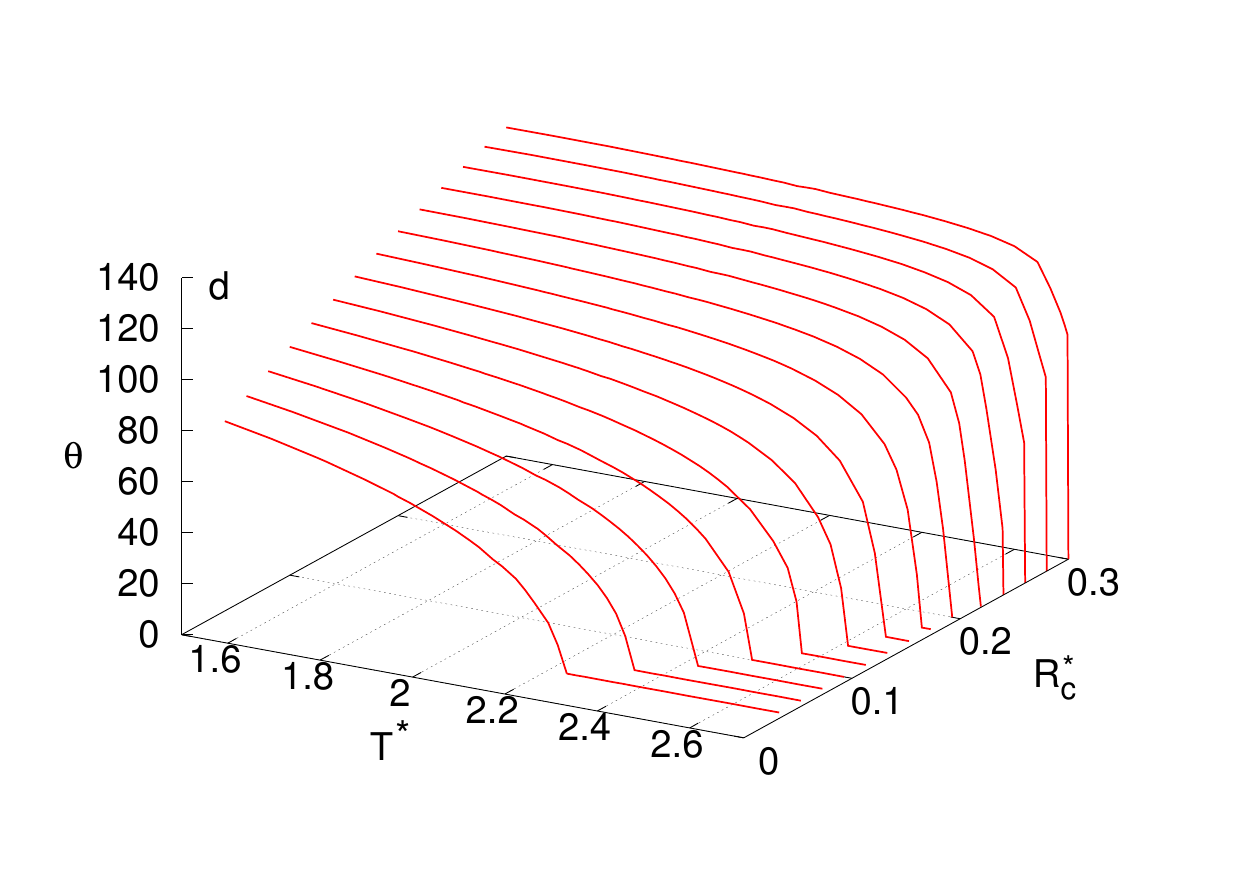}
	\end{center}
	\caption{(Colour online) The dependencies of the static contact angle $\theta$ on temperature and the
		amount of pre-adsorbed particles of the diameter $\sigma_1=0.8$ (part a), $\sigma_1=1$ (part b),
		$\sigma_1=1.2$ (part c) and $\sigma_1=1.4$ (part d). The calculations are for $\varepsilon_{fw}^*=8.311$.} 
	\label{fig:1}
\end{figure}

The first series of calculations were carried out for the surface covered
by a one-component layer of hard-sphere obstacles, i.e., the interaction
of a pre-adsorbed molecule 
with water molecules was of hard-sphere type. Four sizes of pre-adsorbed particles
were studied: $\sigma_1^*=0.8$, 1, 1.2, and 1.4. 

The pre-adsorbed molecules block the access to the surface of 
the adsorbate molecules. As a consequence,
the effective (i.e., averaged over the entire surface) surface-water interactions become weaker.
 Since the wetting of a solid surface results from a balance of 
 solid-fluid and fluid-fluid interactions, one can expect an increase of
 the wetting temperature with the surface density $R_c^*$ of pre-adsorbed species.
 
In figure~\ref{fig:1} we show  three dimensional plots illustrating the dependence
of the contact angle on temperature and on $R_c^*$.  The calculations are
for $\varepsilon_{fw}^*=8.311$.

For $\sigma_1^*=1.4$ (figure~\ref{fig:1}~d), the wetting occurs only for
surface densities $R_c^*\lesssim 0.25$ (or, for $R_c\sigma_1^2 \lesssim 0.49$). 
The close-packed hexagonal surface coverage corresponds 
to the surface density $R_{cp}^*\sigma_1^2=2/\sqrt{3}$.
Therefore,  the covering of approximately 40\% of the 
surface by hard obstacles of the diameter $\sigma_1^*=1.4$  suffices to prohibit the
wetting at all temperatures $T^*<T_c^*$. In the case of the particles of diameter $\sigma_1^*=1$
(figure~\ref{fig:1}~b),
the covering of the surface that inhibits the wetting is higher and equals approximately $R_c\sigma_1^2 \approx 0.82$,
which corresponds to 71\% of the close-packed coverage. 
Of course, the higher efficiency  of larger hard particles in retarding the wetting is
connected with stronger lowering of the effective water-surface interactions by obstacles of larger size.

At low temperatures, the plots of $\theta(T)$ exhibit
a plateau. This type of behavior is more evident for larger obstacles and 
at higher values of $R_c^*$. In other words,  with an increasing $R_c^*$, the surface becomes more hydrophobic for a wider range of temperatures.
We checked that at $T^*=1.5$, the contact angle for the surface covered with the close-packed hexagonal layer
of particles of $\sigma^*=1$ is  $\theta\approx 110^\circ$, while for
$\sigma^*_1=1.4$ it equals $\approx 130^\circ$.

\begin{figure}[h]
\begin{center}
\includegraphics[width=6.95cm]{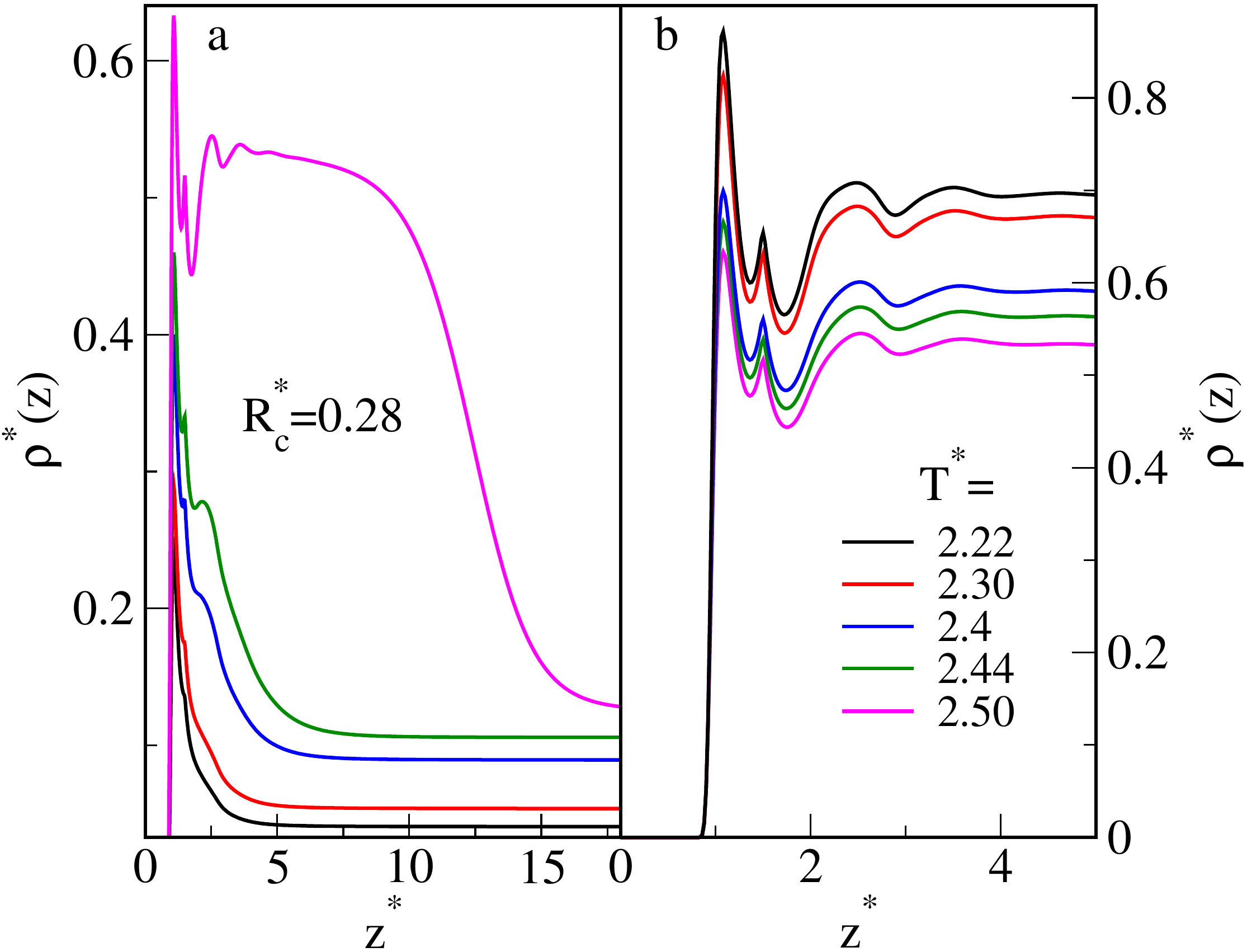}
\includegraphics[width=6.6cm]{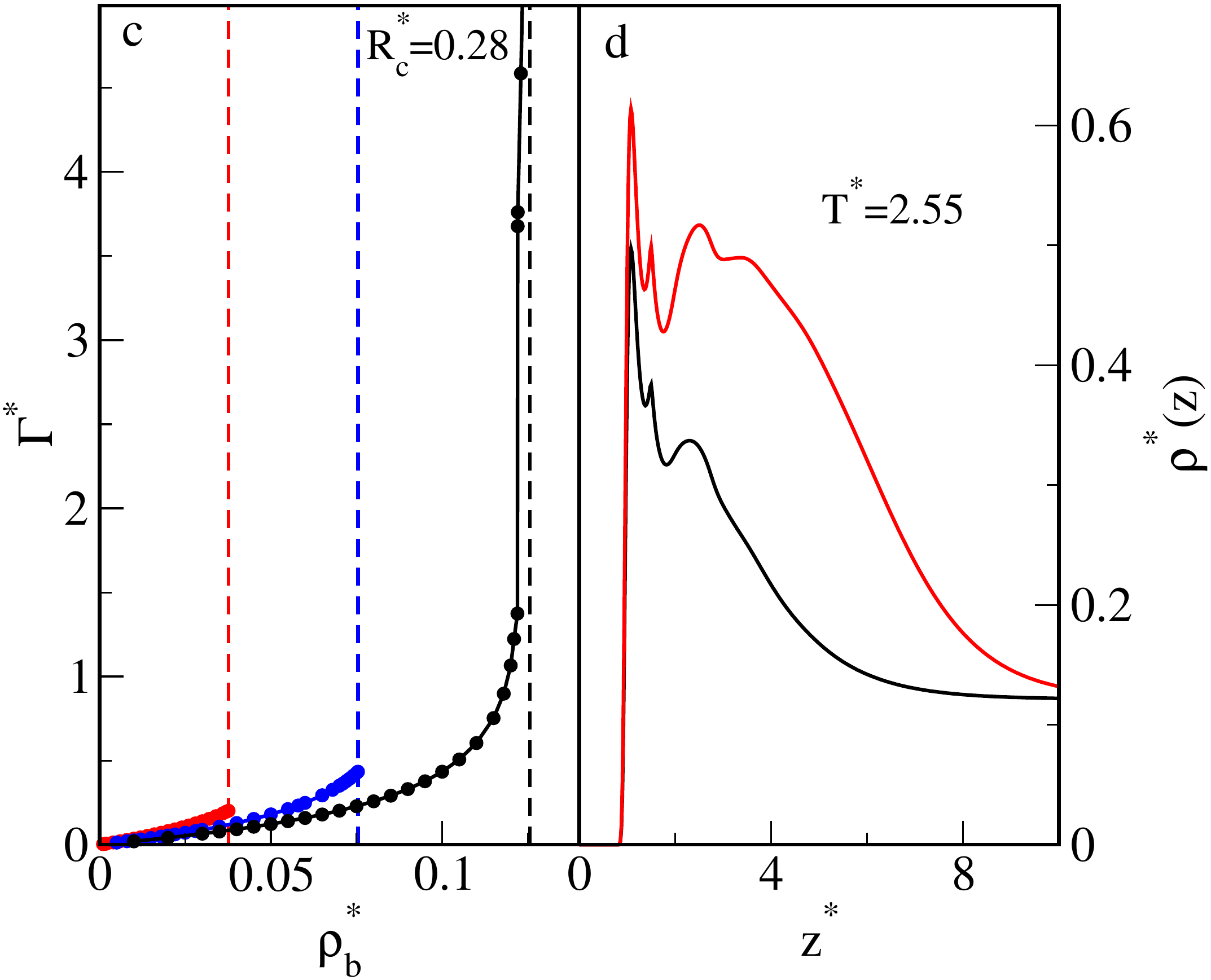}
\caption{(Colour online) Parts a and b. Local density profiles at different
temperatures for bulk density marginally lower than bulk 
dew  density  (part a)
and marginally higher than the bubble
density (part b). The temperatures are given in part b. 
Part c. Adsorption isotherm from  gaseous phase at $T^*=2.2$ (red lines), 2.4 (blue lines), and 2.55 (black lines).
Vertical dashed lines mark the bulk dew densities. 
Part d. Local densities for coexisting thin (black) and thick (red) adsorbed films at $T^*=2.55$
and for the bulk density $\rho^*_b=0.1220$ (the bulk dew density equals $\approx 0.1256$). 
All calculations are 
for $\sigma^*=1$, $R_c^*=0.28$ and for $\varepsilon_{fw}^*=8.311$. }
\end{center}
\label{fig:2}
\end{figure}

Calculations of the contact angles from equation~(\ref{young}) require the knowledge
of the local densities of the systems for bulk densities equal to the densities of 
coexisting liquid and gaseous phases. Figures \ref{fig:2}~a and  \ref{fig:2}~b  show the
density profiles of water at bulk densities marginally lower (part a) and marginally higher
(part b) than the densities at the bulk coexistence. The calculations are for $\sigma_1^*=1$
and $R_c^*=0.28$. Up to the temperature $T_c^*=2.44$ only a thin film formation is observed
for adsorption from the gaseous phase. At $T_c^*=2.5$, however, a thick film is formed
and its thickness diverges as the bulk density approaches $\rho_{bg}^*$. The crossover between
thin and thick film behavior is at $T^*_w=2.445$. At the same temperature, the value of the 
contact angle becomes zero. This indicates a consistency of the values of the wetting
temperature obtained from
the Young equation and from the adsorption study.

\begin{figure}[htb]
	\begin{center}
	\includegraphics[width=6.9cm]{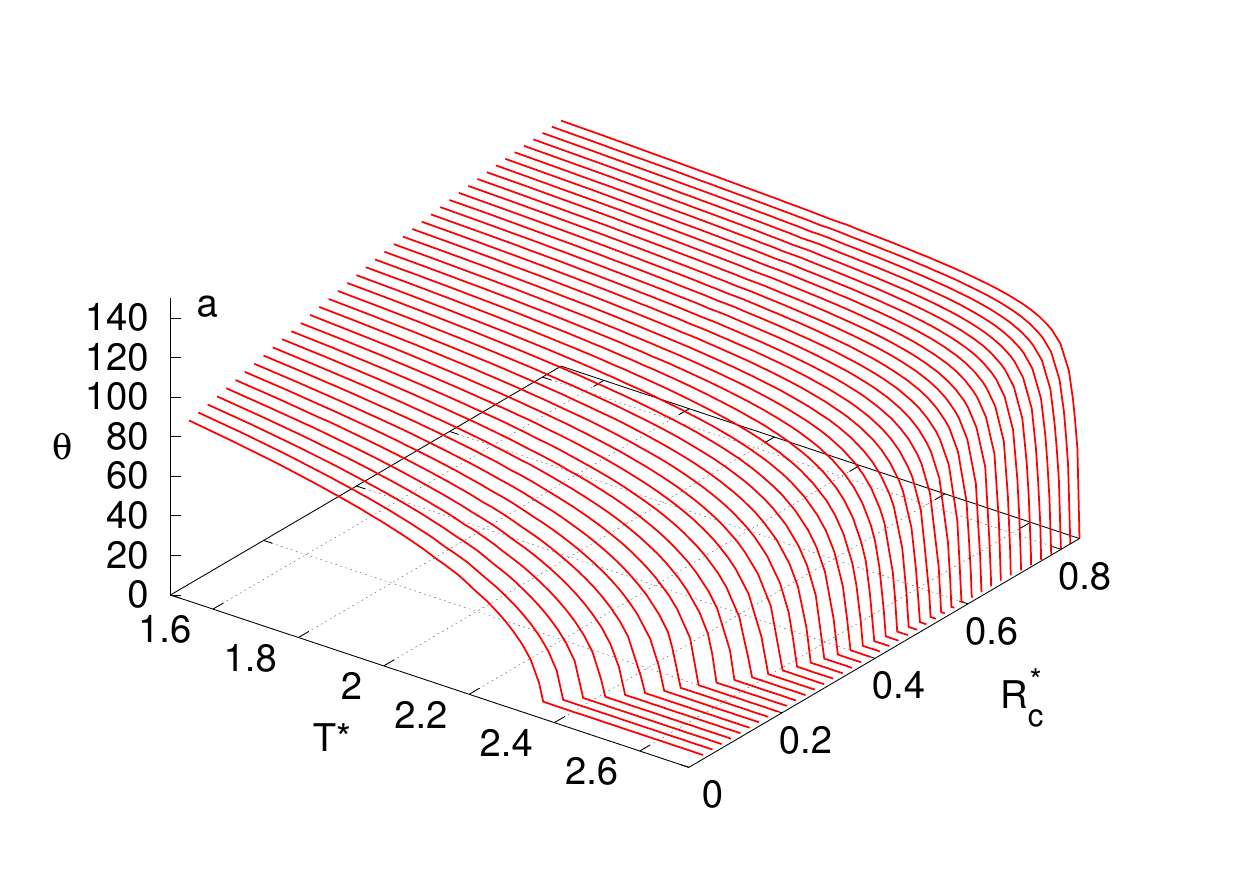}
	\includegraphics[width=6.9cm]{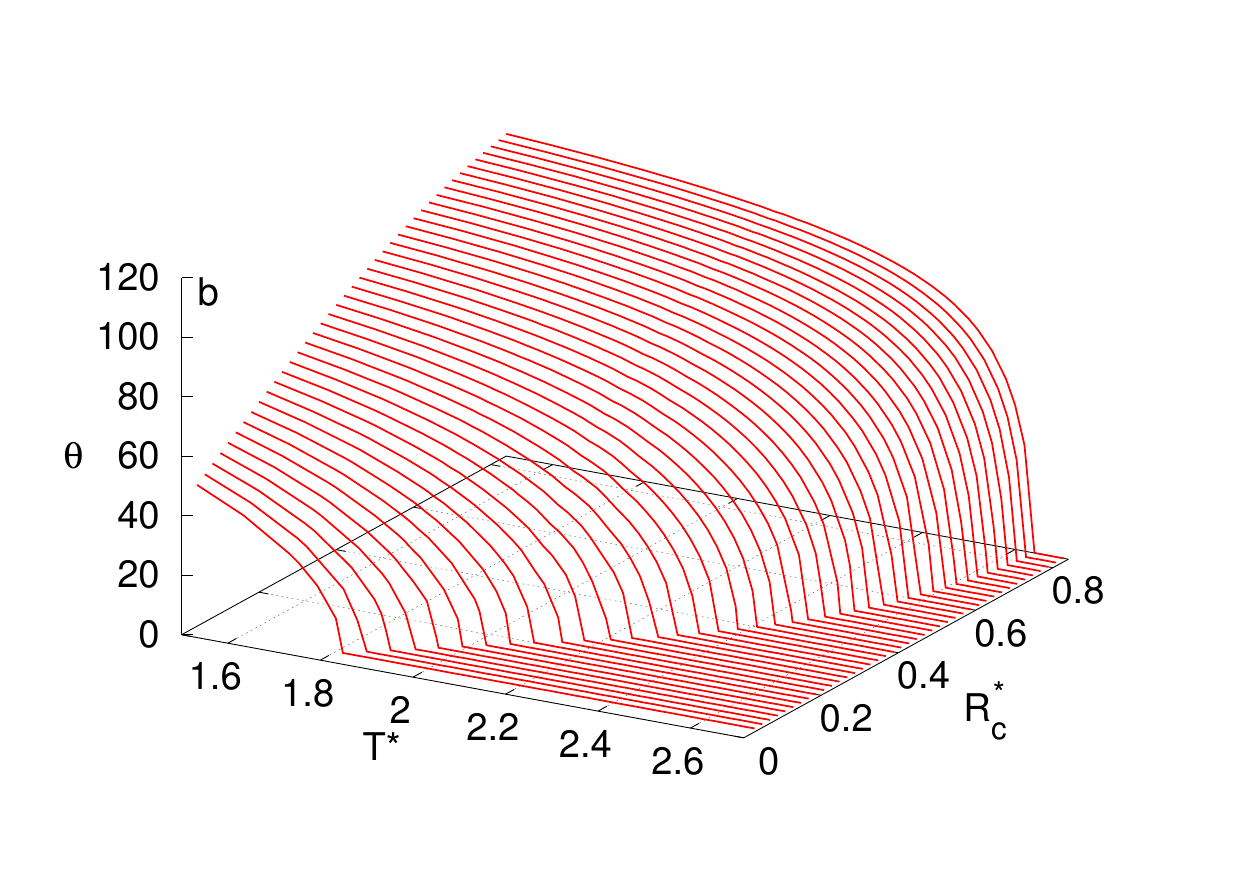}
	\end{center}
	\caption{(Colour online) The dependencies of the static contact angle $\theta$ on temperature and the
		amount of pre-adsorbed particles of the diameter $\sigma_1^*=1$  and 
		for $\varepsilon_{fw}^*=7$
		(part a) and $\varepsilon_{fw}^*=9.5$ (part b).}
	\label{fig:5}
\end{figure}

In figures \ref{fig:2}~c and \ref{fig:2}~d we plot the adsorption isotherms 
at $T^*=2.2$, 2.4  and $T^*=2.55$ (part c), as well as the
density profiles (part d) at $T^*=2.55$. Vertical dashed lines in part c indicate 
the bulk densities of gas coexisting with a liquid. At two lower temperatures, the 
isotherm remains small up to the bulk liquid-vapor coexistence. At $T^*=2.55$, however,
the prewetting jump on the adsorption isotherm appears at $\rho_b^*=0.1220$. The local densities
in part d show the profiles just before and after the prewetting jump. According
to our estimation, for the system under study, the prewetting jump ends at the 
critical prewetting temperature, $T_{cp}^*\approx 2.61$. Although we did not conduct detailed 
studies of the 
prewetting phase transition, the obtained results indicate that for all systems under study,
if the prewetting exists,
it is a first-order phase transition~\cite{evans3,ev,nasza4}.

In the case of non-modified surfaces,
the wetting temperature depends on the depth of the water-surface potential [equation~(\ref{eq:steele})],
$\varepsilon_{fw}^*$. If $\varepsilon_{fw}^*$ increases, the wetting temperature
decreases. A similar behavior was found for surfaces modified
with hard obstacles, which is illustrated in figure~\ref{fig:5}. The calculations
were carried out 
 for $\sigma_1^*=1$.  
Part a shows the dependence
of $\theta$ on temperature and $R_c^*$ for $\varepsilon_{fw}^*=7$, and part b --- for $\varepsilon_{fw}^*=9.5$.
Note that according to the results of our previous work  
for non-modified surfaces 
\cite{nasza4},  the wetting temperature
was $T_w^c=2.29$ for  $\varepsilon_{fw}^*=7$ and $T_w^c=1.76$ for
$\varepsilon_{fw}^*=9.5$.

\begin{figure}[htb]
	\begin{center}
	\includegraphics[width=6.9cm]{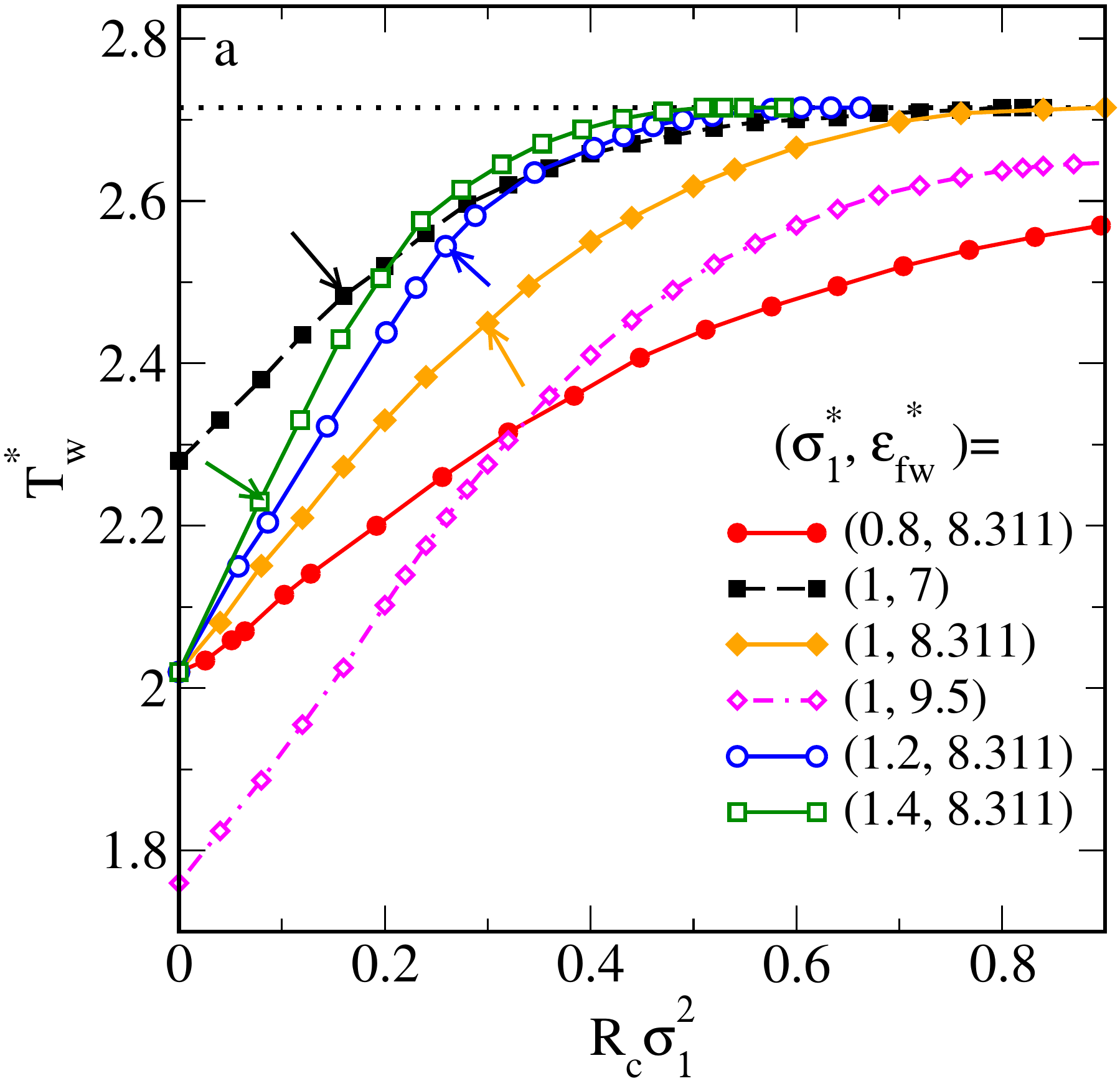}
	\includegraphics[width=6.9cm]{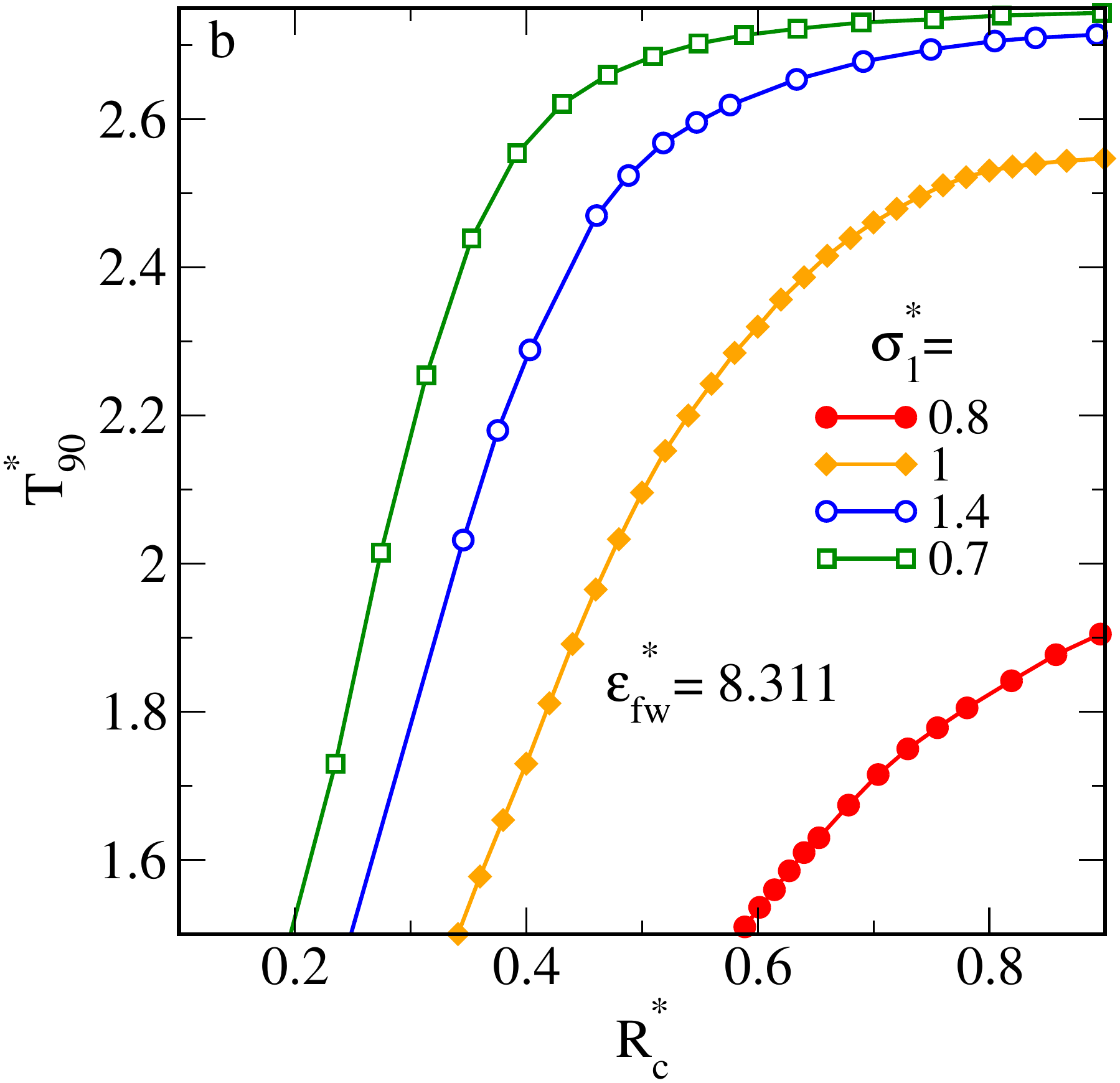}
	\end{center}
	\caption{(Colour online) The dependence of $T_w^*$ (part a) and $T_{90}^*$ (part b) on the
		amount of pre-adsorbed particles for different values of  $\sigma_1^*=1$ and  
		$\varepsilon_{fw}^*$, given in the figure. The dotted line in part a
		denotes the bulk critical temperature.The meaning of arrows is explained in the text.
		We stress that the scale of the abscissa in part a is $R_c\sigma_1^2$, while in part b
		we use the reduced  values of $R_c^*$.}
	\label{fig:6}
\end{figure}

The presence of hard obstacles leads to the effect 
of a further lowering of the effective adsorbing potential
and thus to an increase of the wetting temperature.
 For a weakly adsorbing surface, $\varepsilon_{fw}^*=7$, and
 for high surface density of pre-adsorbed particles, $R_c^*\gtrapprox 0.7$,
  the wetting transition disappears 
for all temperatures $T^* < T_c^*$. 
By contrast, for  $\varepsilon_{fw}^*=9.5$, the wetting transition was observed for
all investigated values of $R_c^*$.

Summary of our calculations for the surfaces covered with hard-sphere obstacles is given
in figure~\ref{fig:6}. Part a presents the dependence of the wetting temperature on $R_c^*$
for different sizes of obstacles and for three values of $\varepsilon_{fw}^*$.
In the experimental studies of wetting, the temperature at which
the contact angle becomes equal to $90^\circ$ is important
since experimental works usually classify
the surfaces with the contact angle $\theta>90^\circ$ as hydrophobic and
the surfaces with  $\theta<90^\circ$ --- as hydrophilic.
The temperature abbreviated as  $T_{90}^*$,
separates these two regimes. We emphasize that in the case of capillaries with hydrophobic
walls, the meniscus of fluid inside the pores is concave, while it is convex for hydrophilic walls. 
The relationship of $T_{90}^*$ on the surface coverage and the size of obstacles is presented
in figure~\ref{fig:6}~b.

Two groups of the curves can be distinguished in figure~\ref{fig:6}~a. The first group  
is at a constant value of
$\varepsilon_{fw}^*=8.311$ and 
illustrates the effect of 
the size of obstacles on the wetting temperature.
All three these curves originate at  ($R_c^*=0, T_c^*=2.02$), i.e., at the wetting temperature
for a non-modified surface.
The second group of the curves is for the fixed value of  $\sigma_1^*=1$ and for three 
values of $\varepsilon_{fw}^*=7$, 8.311 and 9.5. Of course, in this case, the wetting temperature
for different non-modified surfaces is different.

All the curves in figure~\ref{fig:6} exhibit non-linear behavior. Thus their shape contradicts
the predictions by Cassie and Baxter~\cite{cassie} theory. In the case of four curves
(marked by the arrows in figure~\ref{fig:6}~a), there exist the values of the surface coverage 
$R_w\sigma_1^2$ at which the wetting temperature
becomes equal to the bulk critical temperature. For the coverages $R_c\sigma_1^2 > R_w\sigma_1^2$,
the wetting transition is suppressed for all temperatures up to the bulk critical temperature.
The value of $R_w\sigma_1^2$ is lower for higher $\sigma_1^*$ and for lower $\varepsilon_{gs}^*$.
For $\sigma_1^*=0.8$ and $\varepsilon_{fw}^*=8.311$, as well as for $\sigma_1^*=1$ and $\varepsilon_{fw}^*=9.5$,
the wetting transition occurs for all surface coverages $R_c^*$.

\begin{figure}[htb]
\begin{center}
\includegraphics[width=6.9cm]{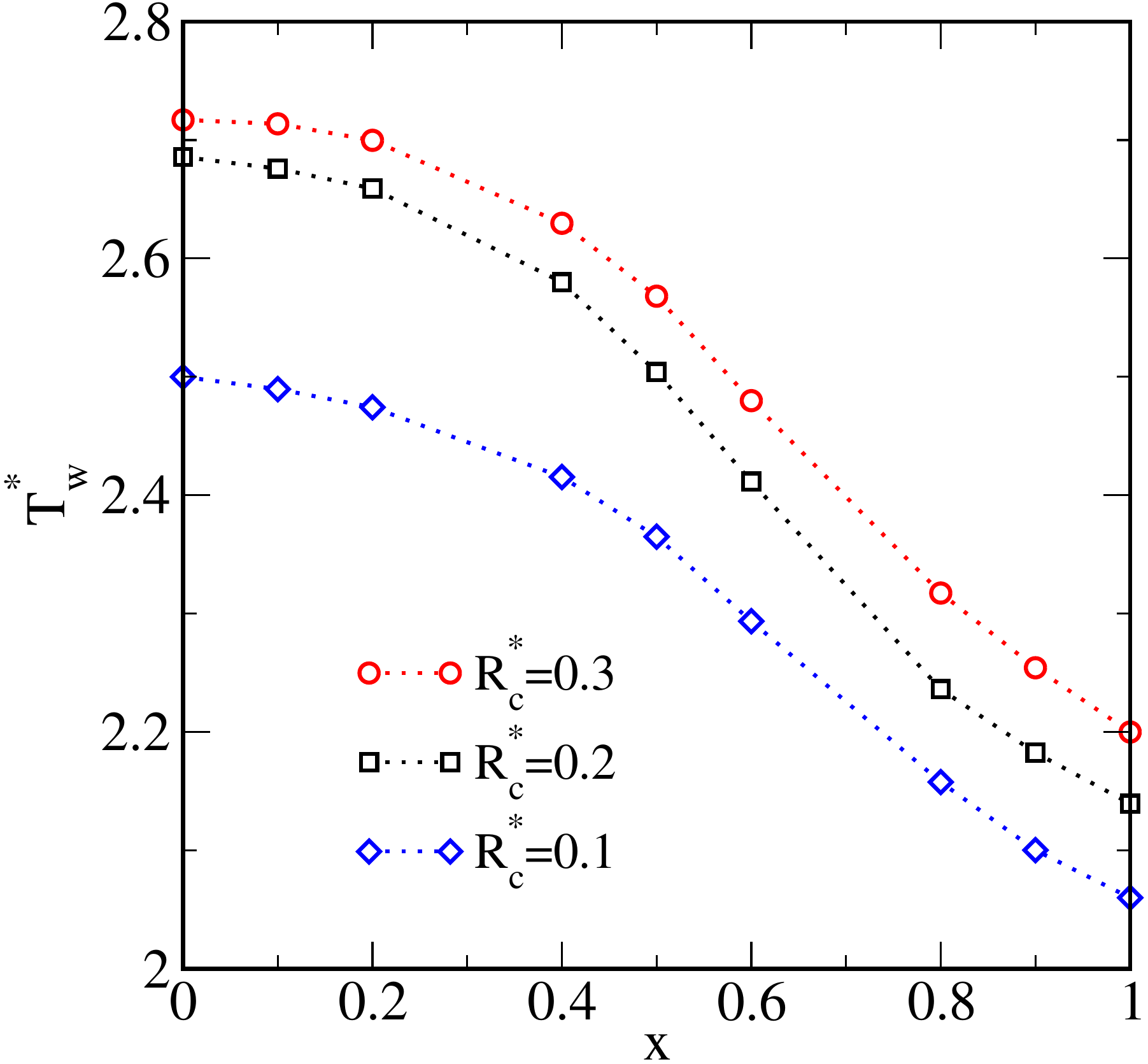}
\end{center}
\caption{(Colour online) The dependencies of $T_w^*$ on the composition $x=R_{c1}^*/(R_{c1}^*+R_{c2}^*)$ of a binary pre-adsorbed phase
of hard-particles of the diameter $\sigma_1^*=0.8$ and $\sigma_2^*=1.4$. The calculations are for fixed
values of $R_c^*=R_{c1}^*+R_{c2}^*$ given in the figure.}
\label{fig:7}
\end{figure}

We also performed the calculation of the contact angles for 
two-component pre-adsorbed layers
consisting
of hard-sphere obstacles of the diameters $\sigma_1^*=0.8$ and  $\sigma_2^*=1.4$. 
Figure~\ref{fig:7} shows the dependence of the wetting temperature on the 
composition of the obstacles, $x=R_{c1}^*/(R_{c1}^*+R_{c2}^*)$.
The displayed results
were obtained at three selected constant values 
of the total two-dimensional
density $R_c^*=R_{c1}^*+R_{c2}^*$. Of course, for $x=0$ the and for $x=1$, the wetting
temperatures are equal to those of one-component layers consisting of spheres
of the diameter $\sigma_1^*=1.4$ and 0.8, respectively. 
Again, contrary to the  Cassie and Baxter~\cite{cassie} theory, the curves
in figure~\ref{fig:7} are non-linear.

The second series of calculations were for a layer of pre-adsorbed molecules
interacting with water molecules via attractive forces. The
interactions between pre-adsorbed particles were still of hard-sphere type.
We assumed the diameter of pre-adsorbed species to be the same as
the diameter of water species, $\sigma_1^*=1$. The energy parameter
of the potential of equation~(\ref{uSW1}) was treated as a  free parameter,
$\varepsilon^*_{f1}=\varepsilon_{f1}/\varepsilon$.

\begin{figure}[htb]
\begin{center}
\includegraphics[width=7.9cm]{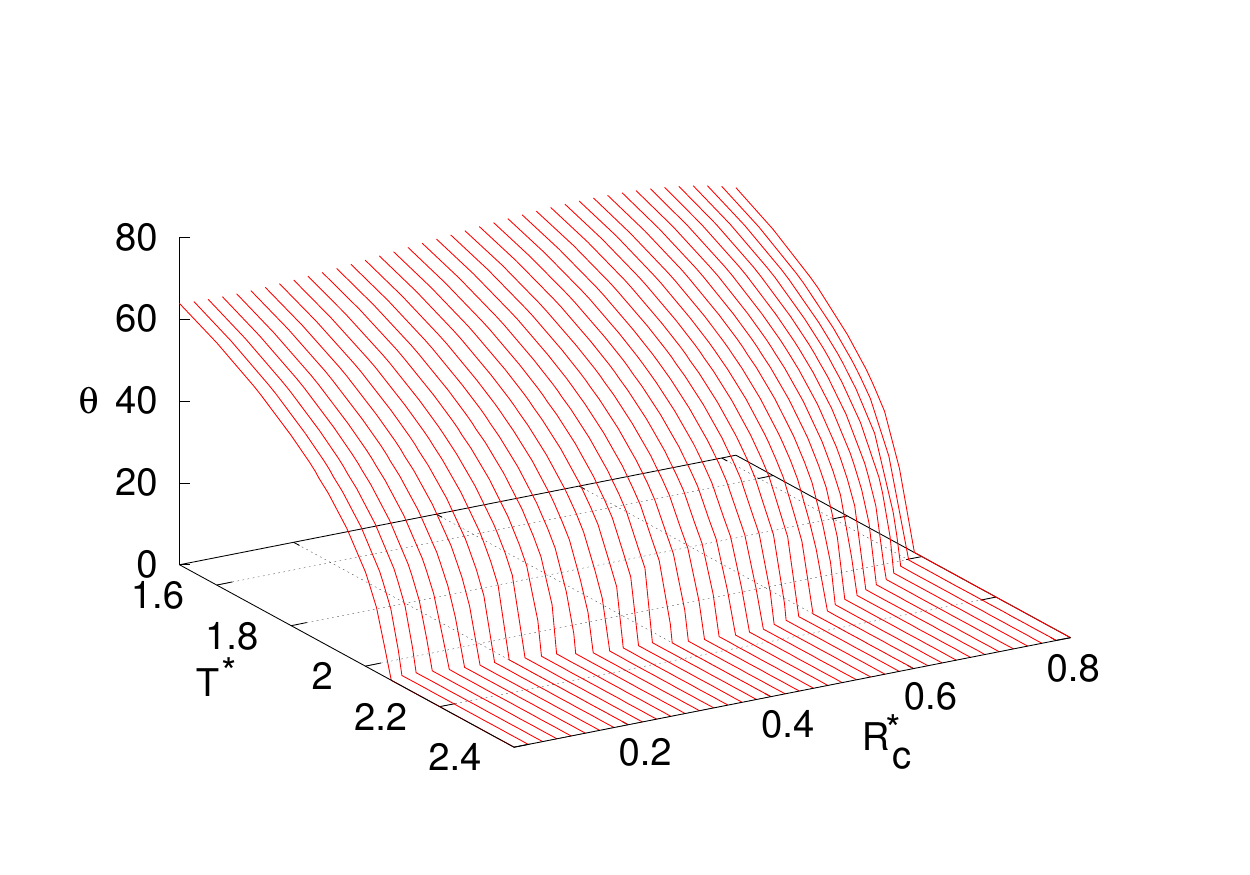}
\end{center}
\caption{(Colour online) The dependence of $\theta$ on $T^*$ and $R_c^*$ for the system with attractive interactions between
water and pre-adsorbed species. Calculations are for $\sigma_1^*=1$ and $\varepsilon_{f1}^*=0.4$.}
\label{fig:att}
\end{figure}

The presence of species attracting the water molecules
leads to two opposite effects. The first one results from
blocking the access of water molecules to the surface. Consequently,
the effective water-surface interactions are lower, and the wetting
temperature  would increase. However, the pre-adsorbed molecules also exert 
attractive forces on water molecules. If the second effect prevails,
the wetting temperature would decrease. Competition between these two factors can lead
to a  
dependence of the wetting temperature on $R_c^*$ 
that could exhibit an extreme.

Figure~\ref{fig:att} shows  a three-dimensional plot of the dependence of
$\theta$ on temperature and $R_c^*$. The presented  results are for 
$\varepsilon_{f1}^*=0.4$. 
At very low values of $R_c^*$, a small decrease 
of the wetting temperature compared to the bare substrate occurs. It means that
for small $R_c^*$, the increase of effective attraction between fluid molecules and
the modified solid is more important than the effect due to the blocking of the surface. 
At still higher
values of $R_c^*$ the wetting temperature starts to increase and attains its maximum value
for $R_c^*\approx 0.51$. Within this region of surface coverages, the
blocking effect plays a dominant role.
A further increase of $R_c^*$ leads to the lowering of the wetting temperature.

\begin{figure}[htb]
\begin{center}
\includegraphics[width=7cm]{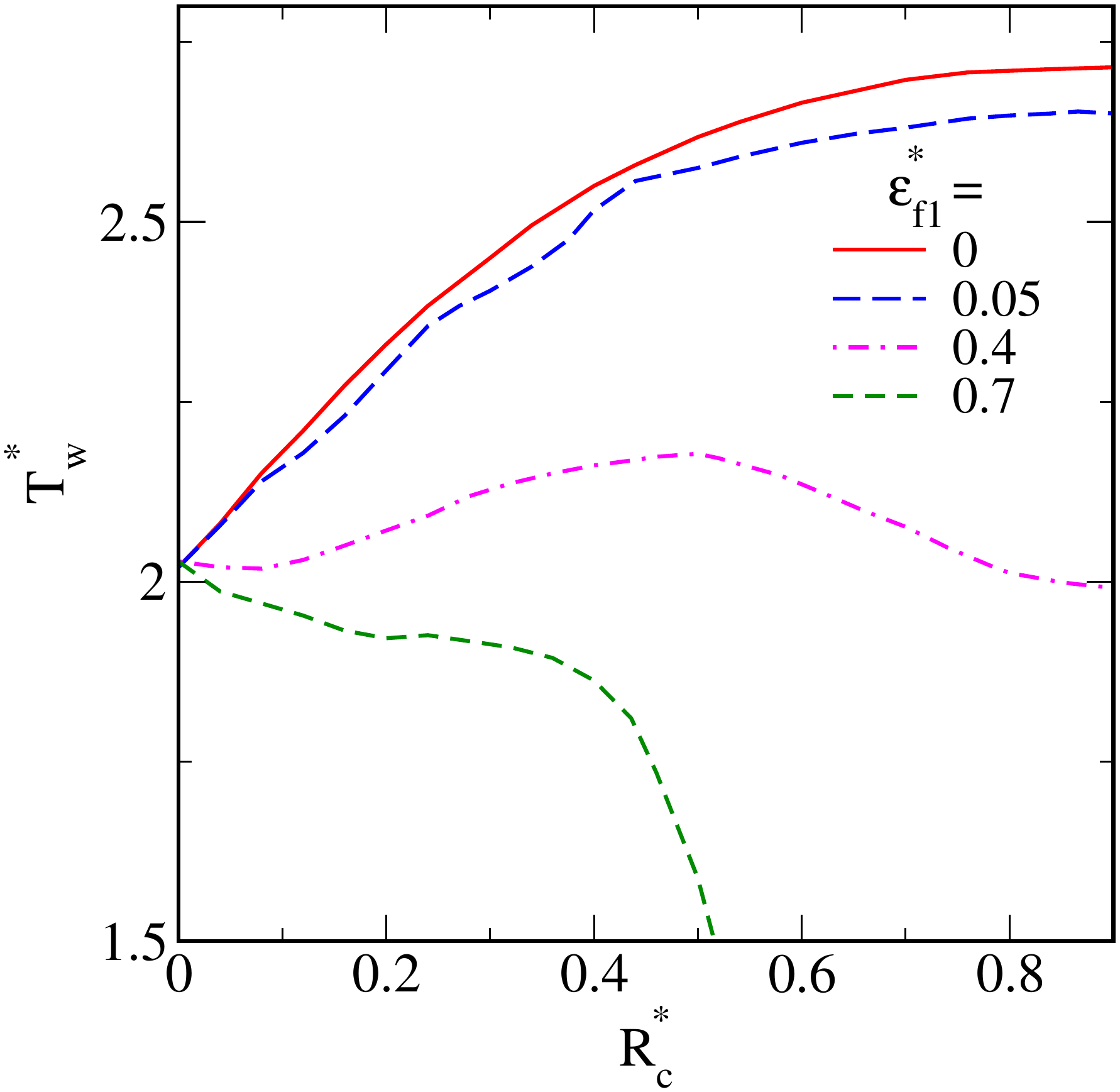}
\end{center}
\caption{(Colour online) The dependence of the wetting temperatures on $R_c^*$ for the system with different values
of $\varepsilon_{f1}^*$ that are given in the figure. The red solid line denotes the results for
pre-adsorbed hard-spheres.
Calculations are for $\sigma_1^*=1$.}
\label{fig:wetaa}
\end{figure}

A comparison of the changes in the wetting temperature with the two-dimensional
the density of pre-adsorbed species and with the energy parameter $\varepsilon_{f1}^*$
is shown in figure~\ref{fig:wetaa}. We also included here the curve 
for the pre-adsorbed layer of hard obstacles as a reference. As expected, for a low value
of $\varepsilon_{f1}^*=0.05$, the evaluated curve $T_w^*(R_c^*)$  is close
to that for hard obstacles. However, we observe an 
undulated course of this function.
The changes in the values of the contact angle result from the 
changes in the free energies for gaseous and liquid phases contacting with a solid.
According to perturbative treatment, the shape of the function $T_c^*(R^*_c)$ is to a great
extent determined by a delicate balance between the hard-sphere contribution to the free energy
and the mean-field term due to water-pre-adsorbed molecule attraction. If the latter contribution
prevails, the wetting temperature decreases, but when the hard-sphere contribution becomes
dominant, the wetting temperature increases.

For $\varepsilon_{f1}^*=0.7$,
the wetting temperature decreases and for all values of $R_c^*$ and for 
$R_c^*>0.4$  this decrease is quite fast.
We did not perform the calculations at temperatures lower than $T_c^*<1.5$, because the 
application of the considered version of the density functional at the temperatures
lower than the triple point temperature may be questioned.

\begin{figure}[h]
	\begin{center}
	\includegraphics[width=7cm]{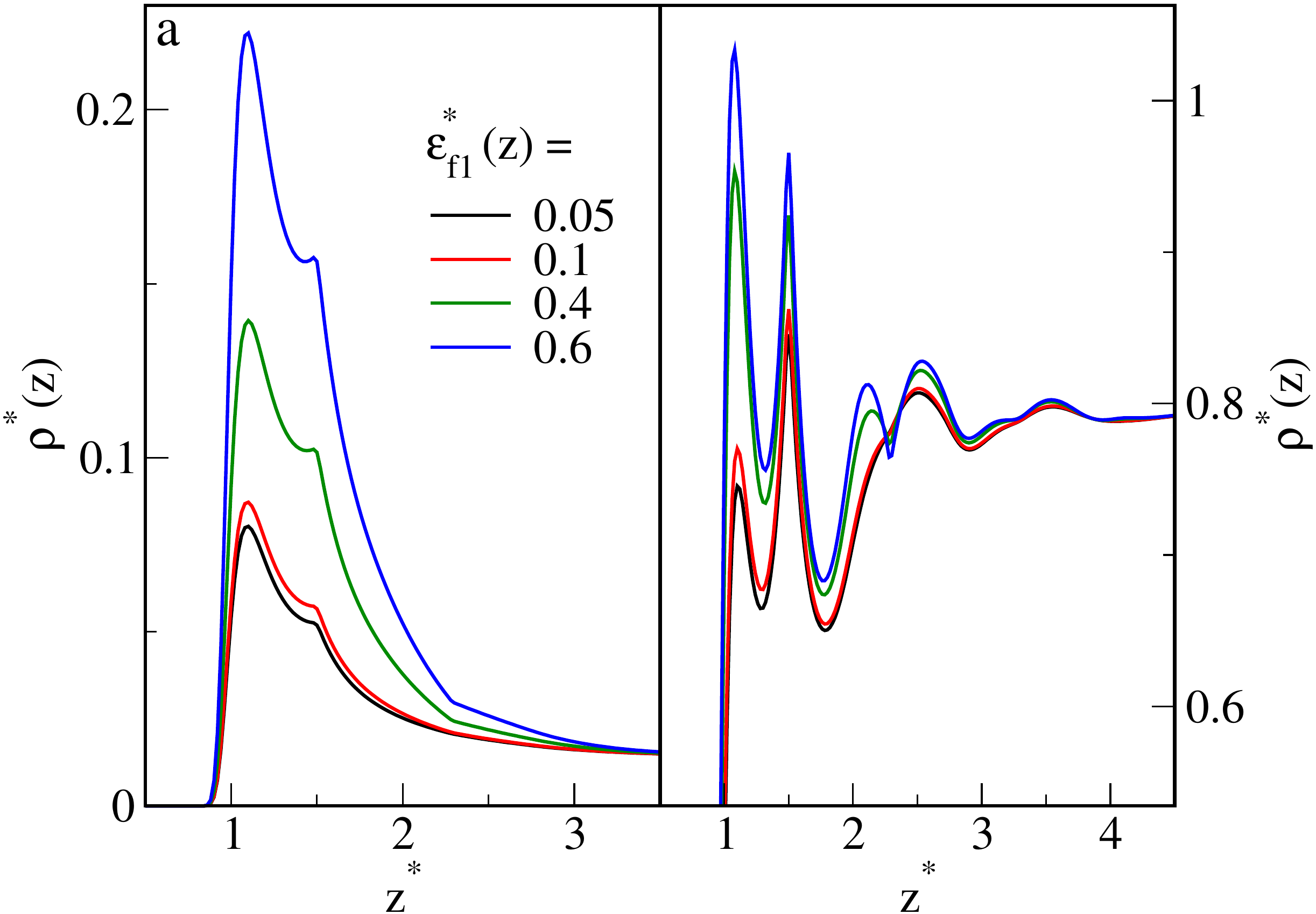}
	\includegraphics[width=7cm]{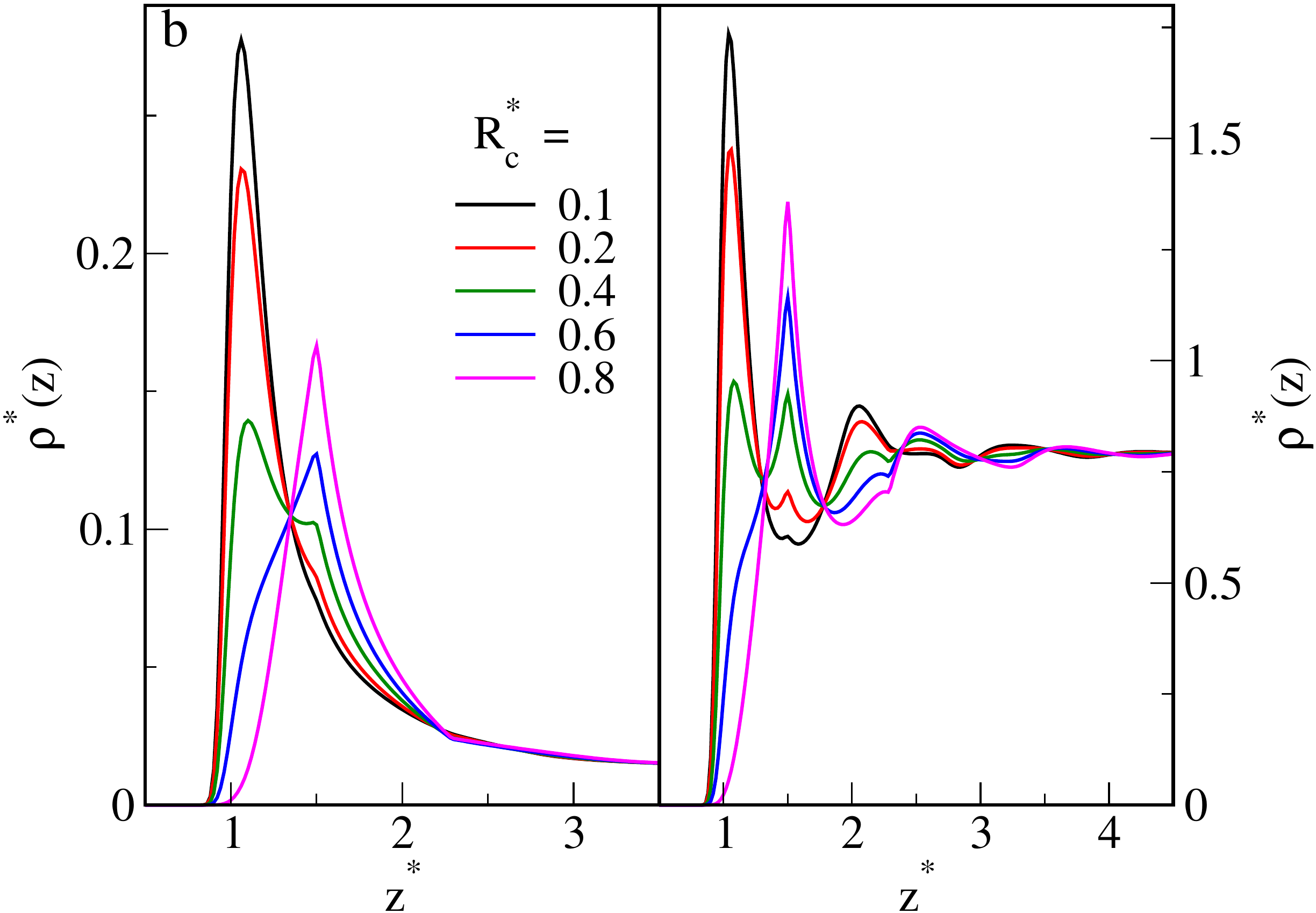}
		\caption{(Colour online) Part a. The dependence of local densities of
			coexisting gaseous (left-hand panel) and liquid (right-hand panel) phases.
			The calculations were carried out at constant $R_c^*=0.4$ and for different
			values of $\varepsilon_{f1}^*$ given in the figure.
			Part b. The same as in part a, but for constant  $\varepsilon_{f1}^*=0.4$ and
			for different values of  $R_c^*$, given in the figure. In all cases
			$\sigma_{1}^*=1$ and $T^*=2$. 
		}
	\label{fig:att_prof}
	\end{center}

\end{figure}

Finally, in figure~\ref{fig:att_prof}
we display the effect of the attractive forces on the structure of gaseous
and liquid water on modified surfaces. Left-hand panels are at bulk densities marginally
lower than the bulk dew density, while right-hand panels are at the bulk liquid
density marginally higher than the bulk bubble density. The presence of sites
attracting water molecules on the graphite surface greately influences
the structure of both gaseous and liquid adsorbed phases. In the case of adsorption
from the gaseous phase, the attraction between pre-adsorbed species and water molecules leads
to the formation of a ``knee'' on the density profile (part a). At higher values of $R_c^*$,
this knee transforms into the second maximum of $\rho^*(z)$. At the highest coverages
of the surface with pre-adsorbed molecules, the local density maximum at the value of
$z$ corresponding to a minimum of the potential $v(z)$ vanishes and the pre-adsorbed layer
starts to play the role of the source of an external potential.

\section{Summary}

We have proposed an approach based on the density functional theory to
describe the changes in the contact angle with the surface heterogeneity. 
According to the proposed model, the surface heterogeneity results
from the formation of a layer of pre-adsorbed molecules on the original, bare 
solid surface.
Depending on the model of interactions between the pre-adsorbed molecules,
on their size, and on their amount (expressed in terms of the surface density,
or surface coverage), different changes in the contact angle and the
wetting temperature were observed. In the 
case of hard-sphere fluid-pre-adsorbed layer interactions,
the presence of pre-adsorbed hard-sphere molecules leads to an increase of
the values of the contact angle and
the wetting temperature. These effects were more pronounced for larger 
pre-adsorbed particles. 

In the case of pre-adsorbed molecules attracting fluid molecules,
the changes in the wetting temperature with the amount of pre-adsorbed species
are more complex. For some selected values of the parameter characterizing the
attractive interaction and
for the selected amount of pre-adsorbed molecules, the wetting temperature can exhibit
a maximum. Since the value of the wetting temperature depends on the surface excess
free energies of gaseous and liquid phases in contact with the modified solid, the 
presence of a maximum is the result of an interplay between particular terms
in the perturbational free energy expansion, basically
on the competitions between the hard-sphere
and the contribution due to attractive water-pre-adsorbed particles 
interaction.

The theory considered in this work indicates that the value of the
wetting temperature does not depend linearly on the surface density of different
kinds of adsorbing sites, as predicted by the classical Cassie approach~\cite{erb}.
Our treatment can also be considered as an alternative to the treatment by Aslyamov et al.~\cite{asly,asly1}. The latter approach is based
on the development of an effective, one-dimensional external potential that, in turn,
is next used in the classical density functional one-dimensional density
functional expressions for evaluating the excess free energy. Instead of evaluating 
the effective free energy, we propose an appropriate modification
of the free energy contributions. Our idea has its origin in an approximate treatment
of the quenched-annealed systems that was previously used to study adsorption
on heterogeneous surfaces~\cite{dft0,dft1,dft2,dft3}. Unfortunately, there exist
no experimental or simulation data that would be useful for verifying our theoretical
predictions.

Basically, our calculations were carried out for one-component
pre-adsorbed phase. However, the theory was proposed for the case of
a multicomponent pre-adsorbed layer. It can be also extended to the case
of a polydisperse mixture by proceeding along the lines described in~\cite{poly,poly1,poly2}.
Next, the theory can be also extended by assuming a multilayer structure of
pre-adsorbed molecules. The treatment of such a system would be based
on the approaches used for quenched-annealed systems~\cite{pizio}.
All these problems are under study in our laboratories.



\ukrainianpart

\title{Кут змочування води на модельній гетерогенній поверхні. Метод функціоналу густини
}

\author{K. Домбровська\refaddr{label1},
	O. Пізіо\refaddr{label2},
	С. Соколовський\refaddr{label1}}

\addresses{
	\addr{label1} 
	Факультет теоретичної хімії, Університет ім. Марії Склодовської-Кюрі, Люблін 20-031, Польща,\\ email: stefan.sokolowski@gmail.com
	\addr{label2}
	Інститут хімії, Національний автономний університет Мехіко,
	Circuito Exterior, 04510, Мехіко, Мексика 
	email: oapizio@gmail.com}

\makeukrtitle
\begin{abstract}
За допомогою методу функціоналу густини розраховано кут змочування у моделі води на гетерогенній графітоподібній поверхні. Неоднорідність поверхні створена шаром преадсорбованих сферичних частинок, або ж сумішшю молекул різних розмірів.
Наявність преадсорбованого шару призводить до виникнен\-ня геометричної та енергетичної гетерогенності поверхонь.
Розглянуто два випадки. Преадсорбовані молекули можуть або поводити себе як перешкоди у вигляді твердих сфер, або можуть також притягувати до себе молекули води. У першому випадку збільшення кількості преадсорбованих складників призводить до зростання температури змочування, яке, однак, не залежить лінійно від кількості перешкод.
У тому випадку, коли перешкоди притягують молекули води, криві, що описують залежність кута змочуван\-ня від кількості преадсорбованих частинок, можуть мати максимум. Крім того, було досліджено, яким чином преадсорбовані складники суміші впливають на локальні густини газоподібних і рідких фаз при їх контакті з модифікованою твердою поверхнею.

\keywords{ метод функціоналу густини, кут змочування, неоднорідна поверхня, модель води}
	
\end{abstract}

\lastpage  
 \end{document}